\documentclass[12pt]{article}

\usepackage[margin=1in]{geometry}
  
\usepackage{algorithm, algpseudocode}

\usepackage[utf8]{inputenc}
\usepackage[pdftex]{graphicx}
\usepackage{afterpage}

\usepackage{amsmath}
\usepackage{breqn}
\usepackage{bm}
\usepackage{subcaption}
\usepackage{comment}
\usepackage{hyperref}
\usepackage{setspace}
\usepackage{threeparttable}
\usepackage[round]{natbib}

\usepackage[title]{appendix}
\usepackage{caption}
\usepackage{amsmath}
\usepackage{booktabs}
\usepackage{lscape}
\renewcommand{\thefootnote}{\arabic{footnote}}
\usepackage{adjustbox}
\usepackage{rotating}
\usepackage{tikz}

\usepackage{float}
\usepackage{footmisc}
\usepackage{amsmath}
\usepackage{amsthm}
\usepackage{enumerate}
\usepackage{subcaption}
\usepackage{adjustbox}
\usepackage[T1]{fontenc}

\usepackage{booktabs} 
\usepackage{graphicx}
\usepackage{tikz}
\usepackage{color}
\usepackage{pgfplots}

\usepackage{cleveref}
\usepgfplotslibrary{dateplot}

\newtheorem{mythm}{Theorem} 
\newtheorem{myprop}[mythm]{Proposition}

\newtheorem{myassump}{Assumption}
\newtheorem{mycor}{Corollary}

\usepackage{caption} 
\usepackage{chngcntr}
\usepackage{apptools}
\AtAppendix{\counterwithin{mylemma}{section}}

\usepackage{bbold}

\newcommand{\ind}{\mathbb{1}}

\newcommand{\ex}[1]{\mathbb{E}\left[#1\right]}

  \title{The Evolution of  Blockchain: from Lit to Dark\thanks{We are grateful to Shihao Yu for valuable comments and feedback.}
}
 \date{}

 \author{
 \large Agostino Capponi\thanks{\scriptsize Corresponding author: Columbia University, Department of Industrial Engineering and Operations Research, 
 Email: \href{mailto:ac3827@columbia.edu}{\texttt{ac3827@columbia.edu}}}, Ruizhe Jia\thanks{\scriptsize  Columbia University, Department of Industrial Engineering and Operations Research, Email: \href{rj2536@columbia.edu}{\texttt{rj2536@columbia.edu}}}, Ye Wang\thanks{\scriptsize ETH Zürich,  Department of Information Technology and Electrical Engineering, Email: \href{wangye@ethz.ch}{\texttt{wangye@ethz.ch}}} 
}
\begin{document}
\vspace{-1cm}
\maketitle
\vspace{-1.2cm}

\begin{abstract}
Transactions submitted through the blockchain peer-to-peer (P2P) network may leak out exploitable information. We study the economic incentives behind the adoption of blockchain dark venues, where users' transactions are observable only by miners on these venues. We show that miners may not fully adopt dark venues to preserve rents extracted from arbitrageurs, hence creating execution risk for users. The dark venue neither eliminates frontrunning risk nor reduces transaction costs. It strictly increases payoff of miners, weakly increases payoff of users, and weakly reduces arbitrageurs' profits.  We provide empirical support for our main implications, and show that they are economically significant. A 1\% increase in the probability of being frontrun raises users' adoption rate of the dark venue by 0.6\%. Arbitrageurs' cost-to-revenue ratio increases by a third with a dark venue.\end{abstract}

\newpage

\section{Introduction}
Blockchain was initially conceived by \cite{bitcoin} as the backbone technology behind digital currencies and decentralized trustless payment systems. Over time, with the development of smart contract technologies, blockchain systems have enabled additional services, such as tokenization of assets, crowdfunding, and decentralized finance (typically abbreviated with {D}e{F}i). See, for instance, \cite{Yermack}, \cite{tokenizationcoporatefinance}, \cite{ICO1} and \cite{introtodf}.

As blockchain evolves from a payment system to an infrastructure for financial services, transparency of information becomes a key concern. Because of the anonymity of the blockchain network, many users cannot send their transactions directly to miners but have to broadcast their transactions through the blockchain peer-to-peer (P2P) network in order to get them executed. Those pending transactions are observable by any node in the network before the execution, including malicious arbitrageurs. Arbitrageurs can exploit information leaked, and execute frontrunning or backrunning attacks on those pending transactions (see, for instance, \cite{anotherAMM}, \cite{frontrunning}). These vulnerabilities are especially pronounced for {D}e{F}i transactions. Arbitrages exploiting pending  {D}e{F}i transactions have generated significant losses for users~\cite{qin2021quantifying,Wang2022impact}, and the losses are often referred to as miner extractable value (MEV\footnote{We refer to \url{https://ethereum.org/en/developers/docs/mev/} for an overview of MEV.}). Moreover, arbitrage transactions make the underlying blockchain more congested, and thus increase transaction costs, which in turn imposes negative externalities on other users of the same blockchain.

Most innovations of blockchain systems have targeted the improvement of the consensus protocol and the system performance. However, few of them have considered the communication mechanism between nodes (especially between users and miners) in the P2P network, which causes the ``built-in'' information leakage problem. 
In the mid of 2021, relay services  such as Flashbots and Eden Network have been introduced\footnote{We refer to \url{https://docs.Flashbots.net/Flashbots-auction/overview/} for an overview of Flashbots relay, and to \cite{Eden} for an overview of the Eden Network.} with the objective of providing protection against frontrunning attacks and mitigating the negative externalities generated from high transactions costs imposed by arbitrageurs on users. Relay services create venues for users to send their transactions directly to miners. We call these venues {\it dark}, because pending transactions submitted through them are not publicly observable, and thus the transaction information cannot be exploited by arbitrageurs.


We build a game theoretical model to analyze the economic incentives behind the adoption of blockchain dark venues. Our model aims at answering the following questions: 
Will the dark venue be adopted by participants of the blockchain ecosystem? Will the adoption reduce frontrunning arbitrage and transaction costs? Is the introduction of a dark venue welfare enhancing? 

We show that the dark venue is at least partially adopted by miners and  utilized by at least one arbitrageur. The introduction of a dark venue neither eliminates frontrunning arbitrage nor reduces transaction costs. It strictly increases the  payoff of miners who adopt the dark venue, but weakly decreases the payoff of miners who stay on the lit venue. After the introduction of the dark venue, the payoff of frontrunnable users increases, while the payoff of arbitrageurs decreases. Aggregate welfare is maximized when all miners adopt the dark venue. However, this outcome may not be attainable in {equilibrium} because miners have a strong incentive to maintain the rents extracted from arbitrageurs. We propose a self-financing {payment} transfer which resolves the misalignment of incentives between miners and users. 

Our model features three types of agents, i.e., miners, users, and arbitrageurs; and two transaction submission venues, i.e., a dark venue (relay) and a {\it lit} venue (the P2P network). Miners need to decide whether or not to join the dark venue. Users submit transactions to the blockchain either through the lit venue or through the dark venue. Transactions sent through the lit venue are publicly observable by all agents, while transactions submitted through the dark venue are observable {only} by miners who join the dark venue. One user faces frontrunning risk when she submits transactions through the lit venue. We refer to her as the frontrunnable user, and to her transaction as a frontrunnable transaction. The remaining users do not face frontrunning risk and are referred to as non-frontrunnable users. Arbitrageurs who identify a frontrunnable transaction in the lit venue compete to exploit the  opportunity.  The adoption rate of miners determines the execution probability in the dark venue. In turn, the venue selection decisions of users and arbitrageurs determine the benefit of joining the dark venue for miners.


Users and arbitrageurs face the trade-off between execution risk and information leakage. On the one hand, using the dark venue alone presents execution risk to users. Transaction submitted to the dark venue face the risk of not being observed by the miner updating the blockchain, who may not have adopted the dark venue. On the other hand, users who only submit through the dark venue avoid the risk of being frontrun. Arbitrageurs who only use the dark venue would not leak out information about the identified opportunity to their competitors. They also gain prioritized execution for their orders because miners on the dark venue prioritize transactions sent through such venue. 
We show that both arbitrageurs and the frontrunnable user will submit their transactions through the dark venue, if sufficiently many miners adopt it. If instead the execution risk is high, arbitrageurs will use both the lit and the dark venue: through the dark venue they gain prioritized execution, and through the lit venue they are guaranteed execution. Because of arbitrageurs' competition, paid transactions fees are above the minimum required for transactions to be executed on the blockchain. {Those fees are passed to miners, and thus miners and arbitrageurs share MEV.}

We investigate the trade-off faced by miners. Each miner can observe more transactions {(i.e., in addition to those submitted to the lit venue)} if he were to join the dark venue. However, if sufficiently many miners join this venue, execution risk becomes sufficiently low to incentivize users to migrate from the lit to the dark venue. This, in turn, eliminates frontrunning arbitrage opportunities that generate MEV. As a result, it may not be incentive compatible for miners to adopt the dark venue.

We characterize the subgame perfect equilibrium of the game. If the frontrunning problem is severe, there exists a unique equilibrium where miners fully adopt the dark venue. The reason is that the frontrunnable user would only submit her transaction through the dark venue, {but} not through the blockchain {lit venue}. In equilibrium, miners fully adopt the dark venue to attract this user and earn the right to observe her transaction. In such case, the incentives of miners and users are {perfectly aligned}. By contrast, if the frontrunning problem is not too severe, there exists an equilibrium where miners do not not fully adopt the dark venue. {The frontrunnable user} would still broadcast through the lit venue and bear the risk of being frontrun. Miners have insufficient incentives to mitigate frontrunning risk because they do not want to forgo MEV. As a result, miners only partially adopt the dark venue and create execution risk, which in turn {leads users to prefer submitting transactions through} the lit venue and be subject to potential frontrunning by arbitrageurs.

In equilibrium, we show that both the aggregate transaction fee per block and the minimum transaction fee required for inclusion in the blockchain increase if a dark venue is present. This may, at first, appear surprising because a dark venue should at least weakly reduce the block space occupied by frontrunning arbitrage orders,  and thus weakly decrease transaction costs. However, this is not the case for the following reasons. First, miners adopt the dark venue if and only if their expected transaction fee revenue increases. Second, the creation of a dark venue raises the number of transactions, because it attracts those which would otherwise not be submitted to the blockchain due to high frontrunning risk. Third, a dark venue increases competition between arbitrageurs and thus raises the bid transaction fees. 

Our analysis generates welfare implications. Miners who join the dark venue are the only participants of the blockchain ecosystem whose welfare strictly increases in the presence of a dark venue. The positional advantage of miners, that is, the ability to determine the execution risk faced by other agents, allows them to extract a larger rent with a dark venue. Welfare of arbitrageurs is reduced because a larger portion of their profits is extracted by miners. {The payoff of users remains unchanged if miners adopt the dark venue only partially to preserve MEV generated from frontrunning arbitrage.} Aggregate welfare of all participants in the ecosystem is maximized if the dark venue is fully adopted by miners. Full adoption eliminates the frontrunning problem, and the entire block space gets allocated to users. However, this outcome is not always attainable in equilibrium because it may not be incentive compatible for miners to fully adopt the dark venue. We propose a self-financing transfer from the frontrunnable user to miners which aligns their incentives. We show that if the frontrunnable user commits to subsidize the dark venue and those subsidies are then passed to miners, the blockchain ecosystem would move to a new full adoption welfare maximizing equilibrium.

We provide empirical support for our model implications. Our dataset contain dark venue transaction-level data of Ethereum blockchain collected from Flashbots API, Ethereum block data, and transaction-level data from Uniswap V2 and Sushiswap AMMs. Our empirical analysis confirms that the dark venue is partially adopted, and further estimates the dark venue adoption rate around $60\%$ as of Jul 2021. Our analysis also shows that miners who join the dark venue have higher revenue than those who stay on the lit venue. Our estimates indicate that joining the dark venue increases miners' revenue by around 0.16 ETH (500 USD) per block.
Consistent with our model prediction that users migrate from the lit to the dark venue if frontrunning risk is large, we find that the probability of being frontrun is positively correlated with the proportion of frontrunnable user transactions submitted through the dark venue. A 1\% increase in the probability of being frontrun increases the proportion of transactions sent through the dark venue by 0.6\%.  Our empirical results also confirm that welfare of arbitrageurs decreases after the introduction of a dark venue. We show that the introduction of dark venue increases arbitrageurs' cost revenue ratio by around a third.

\paragraph{Literature Review.} Our paper contributes to the scarce literature on the information structure of blockchain. \cite{blockchaindisruption} analyze how  blockchain reshapes agents' information and incentives. Our study highlights how different transaction submission channels affect blockchain participants' incentives.

More broadly, our work is related to existing literature on the economic analysis of blockchain systems.  Prior works have studied the economics of consensus protocols~\cite{Folk, PoS, whyprotocalmatters, POSEVOLUATION, permission}, the determination of transaction fees~\cite{transaction1,miningtotrade,chung2021foundations, roughgarden2021transaction}, and mining strategies~\cite{ PoWmining, miningpool, equilibriummining}. We focus on the economic incentives behind the adoption of blockchain dark venues, designed to mitigate the consequences of information leakage. 

Our paper is also related to the branch of market microstructure literature which has analyzed dark venues (e.g., \cite{zhu2014}, \cite{JFEDARKPOOL}, \cite{darkpool3}). These papers study how the introduction of a dark venue impacts market quality and welfare of market participants. \cite{zhu2014} studies how execution risk arising in the dark {venue} leads to better information discovery in the lit venue. 
In his dark pool setting, execution risk arises because informed traders may overcrowd one side of the dark market. In our setting, instead, execution risk in the blockchain dark venue arises because miners earn rents from MEV opportunities, which acts as a disincentive for them to adopt the dark venue.\footnote{In the context of market design adoption, \cite{NBERw25855budish} shows that rent extraction can provide a disincentive for stock exchanges to eliminate sniping risk.}

The paper proceeds as follows. We provide background knowledge of relay services in Section~\ref{instdetails}. We introduce the game theoretical model in Section~\ref{sec:model}. We solve for the subgame perfect equilibrium and examine its economic properties in Section~\ref{sec:SPE}. We analyze welfare implications in Section~ \ref{sec:welfare}. Section~\ref{sec: empirical} provides empirical supports for out model implications. Proofs of technical results are relegated to the Appendix.

\section{Background on Relay Services}\label{instdetails}
In this section, we explain the ``built-in" information leakage problem of blockchain, and discuss the principles of relay services. 

\subsection{Blockchain and Information Leakage}
A blockchain is a decentralized database maintained by distributed participants over a P2P network.
Every participant can issue transactions and broadcast them to every node in the P2P network. Miners, also referred as validators, collect transactions, add them into blocks, and append blocks to the existing blockchain. Users attach an upfront fee to their submitted transactions. Fees allow users to gain execution priority, as miners execute transactions in decreasing order of fees.  

Each node on the blockchain may observe pending transactions in the P2P network. This transparency is not a concern if blockchain is used as a technology for digital payments, because there is no gain to be made from frontrunning a payment transaction. 
Information leakage becomes worrisome if blockchain is used as an infrastructure for financial intermediation. For example, the Ethereum blockchain enables DeFi applications, through which smart contracts act as financial intermediaries and provide  a broad range of financial services, including borrowing and lending, token exchanges, leverage trading, and flash loans. Frontrunning attacks due information leakage can be very costly for users (see \cite{eskandari2019sok}).

Frontrunning attacks include displacement, insertion, and suppression~\cite{torres2021frontrunner}.
In a displacement attack, an attacker observes a profitable transaction from a victim user. She then broadcasts her own profitable transaction with the same arbitrage strategy but with a higher transaction fee. The frontrunning transaction will then be executed in advance of the victim transaction. The attacker will take the profit, while the victim transaction would fail.
In an insertion attack, an attacker observes a frontrunnable transaction from a victim user. She then broadcasts two transactions: one (frontrunning transaction) with a higher transaction fee than the victim transaction and the other (backrunning transaction) with a lower transaction fee. After the frontrunning transaction is completed, the market price changes. Consequently, the price of the victim transaction will be higher than if no attack had taken place. This results in a worse exchange rate and financial losses to the victim, and the attacker receives the profit with the backrunning transaction.
In a suppression attack, an attacker observes an attackable transaction from a victim user. She then broadcasts transactions with a higher transaction fee in order to prevent the victim transaction from being included in the block. Note that the suppression frontrunning attack is very expensive because the attackers try to consume as much gas as possible to reach the capacity limit of the block.
In current DeFi market, {insertion frontrunning attacks are most common.}~\cite{torres2021frontrunner}.

 \subsection{Relay Services}

Relay services are an implementation of the dark venue, which provide a private communication channel between users and miners. 
A centralized relay service receives transactions from users and forwards them directly to  miners, without broadcasting them on the P2P network. 
Therefore, users' transactions cannot be observed by malicious arbitrageurs. To ensure that miners in a private channel do not use observed information, the relay platform screens miners before they join the relay service and monitor their activities.~\footnote{The Flashbots Fair Market Principles (FFMP) can be found at \url{https://hackmd.io/@Flashbots/fair-market-principles}.} 
The first relay service, Flashbots, was launched in January 2021. 

Miners who join the private channel also have to prioritize execution of the highest bidding transactions by including them at the top of a block. The execution order of transactions submitted through the private channel is typically determined by a one-round, seal-bid, first price auction.~\footnote{Flashbots utilizes Coinbase as an additional payment channel between users and miners in addition to the transaction fee attached to the transaction. See \url{https://docs.Flashbots.net/Flashbots-auction/searchers/advanced/coinbase-payment/}} Hence, the transaction submitter neither knows the transactions submitted by other users nor the attached transaction fees. By contrast, in the P2P network, the transaction fee bidding takes the form of an ascending price auction, and it consists of multiple rounds of bid submission. Moreover, pending transactions and their fees are publicly observable.

\section{Model Setup} \label{sec:model}
The timeline of our model consists of three periods indexed by $t$, $t =
 1, 2, 3.$ There are three types of agents: blockchain users, arbitrageurs, and miners. All agents are risk-neutral.

 \paragraph{Miners.} There is a continuum of homogeneous, rational miners. All miners have the same probability of earning the right to append a new block to the blockchain. At the end of period 3, the miner who appends the next block is drawn randomly from a uniform distribution. 
 The winning miner earns the fees attached to the transactions included in the block plus a fixed reward.\footnote{The reward amount does not affect our analysis. Regardless of whether or not a miner adopts the dark venue, his expected block reward remains constant unlike the transaction fees earned.} The miner can at most include $B$ transactions in a block due to limited capacity.

 There exists two transaction submission venues: the lit venue (blockchain P2P network), and the dark venue (relay service).  In period 1, miners decide whether to join the dark venue. We assume that joining this venue is costless for miners.\footnote{As discussed in \url{https://docs.Flashbots.net/Flashbots-auction/miners/faq/}, Flashbots relay is an open-source software and does not charge any fee for usage.}  We denote by $\alpha$ the portion of miners who join the dark venue in period 1.  All miners can observe the transactions submitted through the lit venue, but only miners who join the dark venue can observe the transactions submitted through the dark venue. We assume that miners who join the dark venue do not disclose transaction information.
 
At the end of period 3, the miner who successfully mines the block will select $B$ transactions whose attached fees are the highest. The winning miner can only select from the transactions he observes. We assume that any tie will be broken uniformly at random. The miner decides the execution order as follows. If the miner has joined the dark venue, then he prioritizes the transactions submitted through the dark venue and execute them first.\footnote{Two major relay services, Eden and Flashbots both impose this requirement for miners. See \url{https://docs.Flashbots.net/Flashbots-auction/searchers/faq/.}} Those transactions will be executed in decreasing order of bid fees. Subsequently, the winning miner will include the transactions submitted through the lit venue, {again} in decreasing order of {fees}. A miner who has not joined the dark venue would only include the transactions from the lit venue in decreasing order of fees.

Since a miner's adoption decision does not affect the probability of mining the next block, a miner decides whether to join the dark venue to maximize the expected transaction fees conditional on him successfully mining the next block. The expected transaction fees earned from adopting the dark venue and from using only the lit venue are both contingent on the choice of users and arbitrageurs. We denote the expected fee revenue of the winning miner from adopting the dark venue by $r_{dark}(\cdot)$, and from using the lit venue only by $r_{lit}(\cdot)$.

\paragraph{Users.} {There are two types of users, and the type depends on the exogenously specified nature of their transactions.} 

The first type is a user whose pending transaction is subject to a front-running attack if submitted through the lit venue and identified by arbitrageurs. We refer to this user as frontrunnable and to her transaction as a frontrunnable transaction.  
 If the frontrunnable transaction is successfully written on the blockchain, it generates a benefit $v_0$ to the initiator, i.e., to the frontrunnable user. We assume that $v_0$ is common knowledge. However, if the pending transaction is identified by an arbitrageur, then the arbitrageur can frontrun and earn a profit $c \geq 0$. This, in turn, results in a loss of $c$ for the  frontrunnable user.

 
 The second type of users are those whose transactions are not frontrunnable, even if they are broadcast through the lit venue. We refer to this type of users as the non-frontrunnable users and refer to their transactions as non-frontrunnable  transactions.  Without loss of generality, we assume there exist $B+1$ non-frontrunnable users, indexed by $i \in \{1,2,...,B+1\}$, whose transactions have valuations $v_i, i \in \{1,2,...,B+1\}$ which are common knowledge.\footnote{Having less than $B+1$ transactions would make the analysis of transaction costs trivial, because there would be no competition for block space. We assume that $v_1>v_2>...>v_{B+1}$, and $v_0>v_{B-2}, c>v_{B-2}$.}  We also impose the following technical assumption to rule out corner cases in our analysis:
 
 \begin{myassump}\label{ass:diff}
 The difference $v_{B-2} - v_{B-1}$ is sufficiently small. \end{myassump}

 
 In period 2, users simultaneously decide the venue to which they send their transactions. An user can broadcast her transaction through the lit venue, or through the dark venue, or choose to not submit her transaction. If a frontrunnable transaction is broadcast through the lit venue, it will face the risk of being identified and frontrun by arbitrageurs. If instead a transaction is only broadcast through the dark venue, then it will not be observed by miners who do not adopt the dark venue. Its probability of being included in the next block is at most $\alpha$, which means that the execution risk of the dark venue is determined by miners' dark venue adoption rate. We index the frontrunnable user as user $0$. We denote the channel chosen by user $i, i\in \mathcal{I} = \{0,1,2,...,B+1\}$, by $C_i \in \{\text{Dark}, \text{Lit}, \text{None} \}$. User $i$ also attaches a transaction fee $f_i$ to her transaction. 
 
 
  User $i$ chooses her submission venue $C_i$ and attached fee $f_i$ to maximize her expected payoff: $$U_i = \ex{\ind_{\text{Executed,i}}(v_i -c\ind_{\text{frontrun,i}} -f_i)},$$
 

  where $\ind_{\text{Executed,i}}$ is the indicator function for the event ``transaction by user $i$ is included in the block by miner",  $\ind_{\text{frontrun,i}}$ is the indicator function for the event ``transaction by user $i$ is frontrun by arbitrageurs".
  We assume that users break any tie in favour of the lit venue. Our assumption is justified by the fact that using the dark venue usually requires more sophistication, and the interface for the lit venue is, in general, much easier for users to use.   
 
 
 \paragraph{Arbitrageurs. } There are two competing arbitrageurs, indexed by $j \in \mathcal{J} = \{1,2\}$. The arbitrageurs have to first screen for the pending frontrunnable transaction in the lit venue and then exploit it. An arbitrageur who successfully exploits the opportunity earns a profit $ c \geq 0$. For any pending frontrunnable transaction, each arbitrageur has a probability $p$ of independently {identifying the frontrunning opportunity and exploiting it.} In practice, to identify an arbitrage opportunity and exploit it, an arbitrageur has to screen at least hundreds of pending transactions in a few seconds, calculate the profitability of frontrunning them, construct arbitrage orders, and bid appropriate transaction fees. As a result, not all arbitrage opportunities can be detected and exploited by arbitrageurs, and the probability $p$ captures the difficulty of the above process.


 

 In period 3, two arbitrageurs first search for potential arbitrage opportunities independently. For any exploitable identified  opportunity, the arbitrageur will create an order and decide to which venue to send it:  the lit venue, the dark venue, or both. We assume that if the arbitrageur decides to send an arbitrage order to both venues, then he will give both transactions the same nonce, that is, a unique identifier. Since each nonce can be used only once, at most one of these two transactions will be executed. If the winning miner observes both transactions, he will only include the one with highest transaction fee.  If the order of an arbitrageur is broadcast through the lit venue, the other arbitrageur will observe it and identify the opportunity. The leaked information then leads to more competition for arbitrage execution.\footnote{In practice, arbitrageurs are bots whose addresses do not change often, so their competitors can learn arbitrage opportunities just by tracking the pending transaction submitted from their addresses.} If instead the arbitrage order is only sent to the dark venue, then it may be executed only if the next block is mined by a miner who adopts the venue. Hence, sending arbitrage orders only through the dark venue may limit the probability of the order getting executed, and thus presents execution risk to arbitrageurs.  We then denote the channel chosen by arbitrageur  $j,$ by $V_j\in \{\text{Lit}, \text{Dark},  \text{Both}\}$. We denote the transaction fee bid by arbitrageur $j$ in the private channel by $f_{D_j}$, and in the {lit venue} by $f_{L_j}$. Arbitrageurs choose their strategy to maximize their expected payoff: 
 $$A_j = \ex{\ind_{\text{wins,j}}\ind_{\text{frontrun,0}}(c -f_{executed,j})},$$

 where $\ind_{\text{wins,j}}$ is the indicator function for the event ``{the} order by arbitrageur $j$ is executed before the order by the other arbitrageur", and $f_{executed,j}$ is the transaction fee paid by arbitrageur $j$. 
 
 {Arbitrageurs} employ a mixed strategy when choosing transaction fees. This guarantees the existence of a Nash equilibrium for the subgame in period 3. The tie-break rules for arbitrageurs is that ``both venues" is their first choice, the ``lit venue" is their second choice, and the ``dark venue" is their third choice.

 \paragraph{Transaction Fee Bidding.} The arbitrageur who bids the highest fee can exploit the opportunity.  The transaction fee bidding mechanisms in the two venues are different. Transaction fee bidding in the lit venue is a variant of an English Auction, i.e., an open-outcry ascending-price auction. The auction only has $r$ rounds  where $r$ is a random variable which obeys a geometric distribution with a success rate $\lambda$. There exists a random deadline for the transaction fee bidding auction since the time required for miners to mine the  block is random. In each round, only one arbitrageur moves, and the bid increment has to be larger than $\epsilon$. If only one arbitrageur identifies an opportunity and decides to broadcast his order through the lit venue, then he moves first. If both arbitrageurs identify the same opportunity and decide to send their orders through the lit venue, then the first mover can be either of them with the same probability. To minimize downside risk from the arbitrage execution, arbitrageurs deploy a smart contract. The smart contract would terminate the transaction if the arbitrage opportunity no longer exists. In this case, the transaction would be deemed as failed, and the corresponding transaction fee is negligible and assumed to be equal to zero in our model.  
 
The transaction fee bidding in the dark venue is a one-round, seal-bid, first-price auction, where all bidders only have to submit their bids once to the relay, without leaking any information to other bidders.   If two arbitrageurs submit the same order to exploit the same opportunity, then {only the arbitrageur who pays the highest transaction fee will be considered by miners.}


\paragraph{Equilibrium.} We solve for the subgame perfect equilibrium (SPE) of the game described above. {The equilibrium actions are the dark venue adoption rate of miners $\alpha^*$, the venue selection and transaction fee bidding strategies  of users, and the venue selection and transaction fee bidding strategies  of arbitrageurs.}The strategy of user $i$ is  a mapping from the the dark venue adoption rate of miners, $\alpha$, to her transaction submission venue $C_i$ and transaction fee bid $f_i$. The  strategy  of arbitrageur $j$ is a mapping from the dark venue adoption rate of miners, $\alpha$, and users' actions, $(C_i,f_i)_{i\in \mathcal{I}}$ to his selected venue $V_j$ and transaction fees submitted in each venue $f_{D_j}, f_{L_j} $.

\section{Model Analysis}\label{sec:SPE}
{In this section, we solve for the SPE of the game.} We begin by analyzing the venue choice of arbitrageurs and users. We subsequently study the equilibrium adoption rate of the dark venue, and investigate the corresponding welfare implications. 

\subsection{Venue Choice of Arbitrageurs}
{We analyze arbitrageurs' venue selection strategies, for any dark venue adoption rate $\alpha$ and assuming that the frontrunnable user chooses the lit venue.} Note that it suffices to consider only this choice for the frontrunnable, because if she were to submit her transaction through the dark venue such transaction would not be observable by arbitrageurs. Hence, they would not be able to submit any arbitrage order at $t=3$. 

The main trade-off faced by arbitrageurs is as follows. On one hand, if an arbitrageur chooses only the dark venue, his detected opportunity would not be visible to his competing arbitrageur. This, in turn, reduces competition and thus the arbitrageur's cost from transaction fee bidding. Moreover, the arbitrageur gains prioritized execution, because transactions submitted through the dark venue are placed at the top of the block by miners who join the dark venue. On the other hand, using the dark venue only presents execution risk 
because a fraction of miners may never observe transactions submitted to the dark venue. The following proposition characterizes the choice of the arbitrageurs' venue choice in equilibrium.
 

\begin{myprop}[Venue Selection of Arbitrageurs] \label{venue choice arbitrageurs}
 There exist two critical thresholds $0 < \alpha_1<\alpha_2 \leq 1$, such that:
 \begin{enumerate}

\item If $\alpha \leq  \alpha_1$, then the two arbitrageurs send transactions to both the lit and the dark venues in equilibrium.
\item If $\alpha_1 < \alpha \leq  \alpha_2$, then there are two equilibria. In each equilibrium, one arbitrageur uses both venues while the other arbitrageur only uses the dark venue. 
\item If $\alpha > \alpha_2$, then both arbitrageurs only use the dark venue in equilibrium. 
 \end{enumerate}
\end{myprop}

The main intuition behind the above result is as follows. 
If only a small fraction of miners adopt the dark venue, the execution risk is high. As a result, arbitrageurs will submit their transactions to both venues.
The reason why arbitrageurs would not use only the lit venue is to gain prioritized execution through the dark venue. If instead a large fraction of miners joins the dark venue, execution risk becomes small. The benefit of using the dark venue, that is, of hiding arbitrage opportunities and avoiding intense transaction fee bidding competition, would dominates its cost, that is, execution risk. Hence, arbitrageurs only use the dark venue. The next proposition characterizes the transaction fee bidding strategies of arbitrageurs.

\begin{myprop}[Transaction Fees Bid by Arbitrageurs] \label{strategy arbitrageurs}
 Let $\alpha_1,\alpha_2$ be the critical thresholds identified in Proposition~\ref{venue choice arbitrageurs}. The following statements hold:
 \begin{enumerate}
 
\item If $\alpha \leq  \alpha_1$, then in equilibrium both arbitrageurs bid $c$ in the dark venue. In the lit venue, one of the arbitrageurs places an opening bid  $v_{B-2}$, and afterwards, in each of his bidding rounds, {he} increases  by the minimal increment $\epsilon$ from the previous highest bid.

\item If $\alpha_1 < \alpha \leq  \alpha_2$, then in equilibrium the arbitrageur who uses both venue bids $v_{B-2}$ in the lit venue and  $c$ in the dark venue. The other arbitrageur who only participates in the dark venue bids $c$  if he observes a bid in the lit venue from the other arbitrageur, and bids $v_{B-2}$ otherwise. 

\item If $\alpha > \alpha_2$, then in equilibrium both arbitrageurs  bid a transaction fee $g$ according to the probability distribution
$$P(g)=\left\{\begin{matrix}
\frac{1-p}{p}\cdot\frac{1}{(1-\frac{g-v_{B-2}}{c-v_{B-2}})^2\cdot (c-v_{B-2})} & v_{B-2} \leq g\leq (c-v_{B-2})\cdot p+v_{B-2}\\ 
0 & g >(c-v_{B-2})\cdot p+v_{B-2}
\end{matrix}\right.$$

 \end{enumerate}
\end{myprop}

{If execution risk is high, i.e., $\alpha< \alpha_1$,} arbitrageurs submit their transactions through both venues. Since both arbitrageurs broadcast through the lit venue, if one arbitrageur detects an  opportunity the other arbitrageur will also discover it. Hence, to exploit an opportunity, arbitrageurs have to outbid their competitors. Recall that transactions sent through the dark venue will be prioritized by miners who join this venue. To gain this benefit, both arbitrageurs submit to the dark venue and bid truthfully, that is, bid  transaction fees equal to their profits. In this case, the dark venue induces an arms race for prioritized execution between arbitrageurs. {If execution risk is low, i.e.,
$\alpha > \alpha_2$}, both arbitrageurs use only the dark venue to hide their opportunities. Hence, arbitrageurs do not know whether their competitors have also detected the same opportunity, so the equilibrium must be in mixed strategies. As arbitrageurs no longer bid their {true} valuation in the dark venue, the competition in the dark venue is {less intense relative to the case when execution risk is high.}

{Recall that} if $\alpha_1 < \alpha \leq  \alpha_2$, one arbitrageur only uses the dark venue, while the other arbitrageur uses both the lit and the dark venues. On the one hand, since the latter arbitrageur uses the lit venue, any arbitrage opportunity detected by him will be discovered by the other arbitrageur who uses only the dark venue. This again leads to an arms race for prioritized execution where both arbitrageurs bid truthfully. On the other hand, any arbitrage opportunity detected by the arbitrageur who uses only the dark venue will not be visible to the other arbitrageur. Hence, there will not be any competition, and the arbitrageur who uses only the dark venue can bid the minimum transaction fee.   

Observe that the transaction fee paid by arbitrageurs is pocketed by the winning miners. Because of competition, the transaction fees bid by arbitrageurs are always higher than $v_{B-2}$, that is, the minimum fee which guarantees a transaction to be executed by miners. This suggests that  miners extract a portion of MEV.  

\subsection{Venue Choice of Users}
We analyze the venue selection strategy of the frontrunnable user, for an exogenously specified relay adoption rate $\alpha$. 

The main trade-off faced by the frontrunnable user is  straightforward. Using the dark venue exposes her to execution risk but eliminates the risk of being frontrun. {Unlike arbitrageurs, the frontrunnable user does not use the dark venue to outbid competitors but merely to avoid frontrunning.} When the dark venue adoption rate of miners is sufficiently large, the execution risk is small, and then the user will also adopt it to avoid frontrunning.

 The following proposition characterizes her strategy in equilibrium:

\begin{myprop}[Venue Selection of Users] \label{users proposition}
 There exist three critical thresholds $0<\lambda_1< \lambda_2< \lambda_3<1$ such that the frontrunnable user sends {her} transaction through the dark venue:
 \begin{enumerate}
     \item  If and only if $\alpha > \lambda_1$ whenever $\alpha \in [0,\alpha_1]$.
     \item  if and only if $\alpha > \lambda_2$ whenever $\alpha \in (\alpha_1, \alpha_2]$.
     \item  if and only if $\alpha > \lambda_3$  whenever $\alpha \in (\alpha_2, 1]$. 
 \end{enumerate}
\end{myprop}

{The thresholds for adoption of the dark venue by the arbitrageurs and by the users depend on the probability $p$ that an arbitrageur detects the opportunity. The following corollary characterizes how these thresholds vary with $p$, keeping every other parameter fixed.}

\begin{mycor}[Sensitivity Analysis] {The signs of the sensitivities of $\alpha$'s and $\lambda$'s with respect to $p$ are as follows:}
\begin{enumerate}
    \item $\frac{\partial \lambda_1}{\partial p} < 0, \frac{\partial \lambda_2}{\partial p} < 0, \frac{\partial \lambda_3}{\partial p} < 0$ 
    \item $\frac{\partial \alpha_1}{\partial p} > 0, \frac{\partial \alpha_2}{\partial p} > 0$ 
\end{enumerate}

\end{mycor}
As $p$ increases, the risk of being frontrun increases, and thus the benefit of using the dark venue for the frontrunnable user increases. Hence, threshold for the adoption of the dark venue  decreases. {Vice-versa, as $p$ increases it becomes easier to detect an arbitrage opportunity, reducing the value of information about the arbitrage opportunity. Hence, arbitrageurs are less incentivized to use the dark venue for protecting their private information.}

    
    
    
    
    
    
    
    
    
    
    

\subsection{Miners' adoption and Equilibrium}
We derive the equilibrium dark venue adoption rate of miners, $\alpha^*$, {and characterize the SPE.} 

For any $\alpha > 0$, the miners who join the dark venue receive a higher payoff than those who only stay in the lit venue: 
$$r_{dark}(\alpha) \geq r_{lit}(\alpha).$$
This is because transactions submitted through the dark venue can only be observed by miners who adopt the dark. As a result, if the actions of users and arbitrageurs are fixed, each individual miner has an incentive to join the dark venue. 

{The situation changes once we account for the strategic responses of users and arbitrageurs. If sufficiently} many miners join the dark venue, that is, if $\alpha$ is large enough, then the payoff of each miner may be lower than their payoff when $\alpha = 0$. This is because the frontrunnable user may then route her transaction from the lit to the dark venue if the execution risk in the dark venue is  small enough. The migration of this transaction would eliminate frontrunning opportunities {and thus reduce} MEV. 

We first characterize the equilibrium strategy of the frontrunnable user in the benchmark case where there is no dark venue. This is obtained from our game theoretical framework by setting  $\alpha = 0$, and considering the subgame at periods $t=2,3$. 

\begin{myprop}[Only Lit Venue Benchmark]\label{equilibrium: benchmark}
When $\alpha=0$, there exists a threshold $c_1 \geq 0$ such that the frontrunnable user submits the transaction to the blockchain if and only if $c \leq c_1$.
\end{myprop}

If the frontrunning problem is severe, i.e., $c>c_1$, then the frontrunnable user is not willing to submit her transaction to the blockchain because the cost of being frontrun exceeds the benefit of executing her transaction. Conversely, if the frontrunning problem is not too severe, i.e., $c \leq c_1$, then the frontrunnable user submits to the blockchain even if she faces the risk of being frontrun. 

We next characterize the SPE of our model. We refer to the equilibrium where the relay adoption rate $\alpha^* =1$ as the \textit{full adoption equilibrium},  the equilibrium where the relay adoption rate $ \alpha^* \in (0,1)$ as the \textit{partial adoption equilibrium}, and the equilibrium where the relay adoption rate $ \alpha^*  = 0$ as \textit{no adoption equilibrium}. 

\begin{myprop}[Characterization of the Equilibrium ]\label{equilibrium: from lit to dark}
Let $c_1$ be the critical threshold identified in Proposition~\ref{equilibrium: benchmark}. The following statements hold for the SPE of the game:
\begin{enumerate}
    \item If $c > c_1$, there exists a unique full adoption equilibrium where the relay adoption rate $\alpha^* =1$, the frontrunnable user selects the dark venue, and the arbitrageurs do not submit arbitrage orders. 
    \item If $c \leq c_1$, there exists a partial adoption equilibrium where the relay adoption rate $\alpha^* < 1$, the frontrunnable user submits her transaction through the lit venue, and the arbitrageurs send their orders to the dark venue only or to both venues.
\end{enumerate}
\end{myprop}
The dark venue will be, at least partially, adopted by miners, and the equilibrium outcome is contingent on the severity of the front-running problem.  Suppose the frontrunning problem is severe.
In the absence of a dark venue, it is too costly for the frontrunnable user to submit transactions to the blockchain. To incentivize the frontrunnable user to submit and earn the transaction fee, miners adopt the dark venue. In equilibrium, all miners decide to join the dark venue so that they are able to observe the transaction submitted by the frontrunnable user. 

Suppose the frontrunning problem is not too severe. Even without a dark venue, the frontrunnable user would still submit her transaction to the blockchain even if she bears the risk of being frontrun. Frontrunning arbitrage generates MEV for miners. To maintain their MEV, only a small fraction of miners choose to adopt the dark venue, which creates high execution risk. As a result, the frontrunnable user prefers to submit through the lit venue and face frontrunning risk. In such case, the introduction of a dark venue does not prevent frontrunning arbitrage.  

\section{Welfare Implications}\label{sec:welfare}
We investigate how the introduction of a dark venue impacts transaction costs on blockchain. We also analyze how  welfare of market participants is impacted by a dark venue. 


 We impose the following equilibrium selection criterion. Among all equilibria characterized in part 3 of Proposition \ref{equilibrium: from lit to dark}, we select the equilibrium which maximizes the aggregate payoff of all miners. We pick this specific equilibrium, because it is the most likely to be coordinated upon by miners. Big mining pools can coordinate and move a sufficiently large mass of mining power from one venue to the other. Hence, they can always achieve the equilibrium that maximizes their aggregate payoff. We also remark that our result is robust to our equilibrium selection, and selecting any partial adoption equilibrium in part 3 of Proposition \ref{equilibrium: from lit to dark} will yield the same results on welfare. 
 

\subsection{Transaction Costs on Blockchain}
We begin by showing that the introduction of a dark venue does not serve its intended purpose of reducing blockchain congestion and transaction costs.

\begin{myprop}[{Transaction Costs with Dark and Lit Venues}]\label{transaction cost}
The introduction of {a dark venue} increases both the total fee of all transactions in a block and the minimum fee that guarantees the execution of a transaction. 
\end{myprop}

Because the introduction of a dark venue weakly reduces the block space used by arbitrageurs, one would expect a decline in  transaction costs. Our analysis shows that this is not the case for the reasons outlined next. Miners would adopt the dark venue only if they earn higher transaction fees, and thus the equilibrium transaction costs increase. Moreover, as shown in part 2 of Proposition \ref{equilibrium: from lit to dark}, the introduction of the dark venue may generate more frontrunnable transactions on the blockchain. This drives up the demand for block space and consequently results in higher transaction costs. Further, as shown in Proposition \ref{strategy arbitrageurs}, the introduction of the dark venue may strengthen the arms race for {first} execution between arbitrageurs. The latter would bid high transaction fees in the dark venue, which leads to a higher MEV and transaction costs. 

This result implies that the negative externality induced by MEV cannot be mitigated by the introduction of a dark venue, because it is not incentive-compatible for miners to give up their rents extracted from users and arbitrageurs.


\subsection{Welfare Analysis}
We study how the introduction of a dark venue affects the welfare of the agents in the blockchain ecosystem as well as the aggregate welfare. 

\begin{myprop}[{Welfare of miners, user, and arbitrageur}] \label{welfare analysis}
The introduction of the dark venue leads to
\begin{enumerate}
    \item a strict increase in  aggregate welfare of miners,
    \item a strict increase in welfare for miners who adopt the dark venue, and a decrease in welfare for miners who do not adopt the dark venue,
    \item an increase in welfare for the frontrunnable user,
    \item a reduction in welfare for arbitrageurs. 
\end{enumerate}
\end{myprop}

Miners' increase in welfare can be decomposed into two parts: an increase in MEV, and an increase in transaction fees due to a higher demand for block space. First, recall from Proposition \ref{strategy arbitrageurs} that the introduction of the dark venue exacerbates competition between arbitrageurs and increases  MEV. This, in turn, leads to a reduction in welfare for arbitrageurs, because a higher portion of their arbitrage profits is transferred to miners. Second, recall that the presence of a dark venue may incentivize the frontrunnable user to submit her transaction to the blockchain and thus increase the demand for block space. This, in turn, increases miners' revenue from transaction costs. Even though the aggregate welfare of miners increase after the introduction of the dark venue, the {welfare} of miners who do not join the dark venue weakly decreases.  This is because some transactions migrate from the lit to the dark venue, and miners who stay in the lit venue can no longer observe them. 

The welfare of the frontrunnable user increases because she has now access to a privacy-preserving transaction submission venue. It is worth observing that her welfare does not necessarily increases strictly. Unless the frontrunning problem is very severe, miners adopt the dark venue partially and create execution risk. As a result, the frontrunnable user may find it preferable to stay in the lit venue and bear frontrunning risk.



We next analyze aggregate welfare, defined as the sum of expected payoffs of miners, users, and arbitrageurs. 

\begin{myprop}[Aggregate Welfare] \label{aggregate welfare}
The followings statements hold:
\begin{enumerate}
    \item The aggregate welfare is maximized when the dark venue is fully adopted by miners. 
    \item  The introduction of the dark venue weakly raises aggregate welfare.
    \item If $c>c_1$, then the unique full adoption equilibrium attains the maximum aggregate welfare; if $c\leq c_1$, then any partial adoption equilibrium yields an aggregate welfare strictly below the maximum.
    

\end{enumerate}
\end{myprop}

The above result can be intuitively understood as follows. The profit of arbitrageurs and fee revenue of miners are merely transfers of wealth from users. Despite MEV is extracted from arbitrageurs by miners, it is just a fraction of the profits arbitrageurs extracted from users. 
As a result, aggregate welfare is maximized if the sum of the valuation of users' transactions added to the block is maximized.  In particular, {maximum} welfare can only be achieved if frontrunning arbitrage does not take up any block space. If the dark venue is fully adopted by miners, execution risk is small, and the frontrunnable user submits through the dark venue. Because no arbitrageur demands for block space, the block only includes the $B$ users' transactions with the highest valuations, and the aggregate welfare is then maximized. 


{We have shown that the introduction of the dark venue weakly improves aggregate welfare. Moreover, the private and social optimum coincide if the frontrunning problem is severe.}
However, if the frontrunning problem is not too severe, the ecosystem would coordinate on a partial adoption equilibrium where frontrunning arbitrage is still present, and the block space allocation would not be efficient. The aggregate welfare maximizing outcome is then unattainable because miners have a positional advantage and can determine other participants' execution risk. For miners, the dark venue merely serves to extract larger rents. 
 
We propose a self-financing transfer from the frontrunnable user to miners so that the misalignment of incentives is resolved, and the resulting full adoption equilibrium achieves the welfare maximizing outcome.
 \begin{myprop}[Attaining Full Adoption]\label{transfer design}
  There exists $\theta \geq 0$ such that if the frontrunnable user commits at $t=1$ to make a payment $\theta$ to the winning miner on the dark venue, then 
  (i) a unique full adoption equilibrium is attained; (ii) the expected payoff of all miners strictly increase;  (iii) the expected payoff of the frontrunnable user does not decrease. 
 \end{myprop}
 
In the partial adoption equilibrium, the MEV earned by miners is only a fraction of arbitrageurs'  profit extracted from the frontrunnable user. If the frontrunnable user commits to  make a payment to the winning miner on the dark venue, and this payment is above miners' expected MEV in the partial adoption equilibrium, then  it is incentive compatible for all miners to adopt the dark venue, and the aggregate welfare is maximized. The payoff of the frontrunnable user in the full adoption equilibrium net of the payment is strictly higher than her payoff in the partial adoption equilibrium (where no transfer between the user and miners occurs).   This proposed transfer could be implemented in a straightforward manner. The relay service can set up a reward pool which allows users to voluntarily deposit ERC-20 tokens into it. Any miner who joins the relay service and successfully mines a new block that includes transactions sent through the relay service can claim the tokens deposited in the reward pool.

\section{Empirical Analysis} \label{sec: empirical}
In this section, we provide empirical support to the implications of our model. Section \ref{sec implications} lists the model implications we validate. Section \ref{sec data} describes our dataset. Section \ref{sec stylized facts} defines the key variables and stylized facts. Section \ref{sec empirical results list} describes our empirical results. 

\subsection{Testable Implications} \label{sec implications}
Our model generate the following implications:

\begin{enumerate}
    \item The blockchain dark venue will be partially adopted by miners (see  Proposition~\ref{equilibrium: from lit to dark}). 
    \item Miners who adopt dark venue have a higher expected payoff than miners who stay in the lit venue. (See part 1 of Proposition~\ref{welfare analysis})
    \item Users submit transaction through the dark venue when the frontrunning risk is high (see Proposition~\ref{equilibrium: from lit to dark}).  
    \item Arbitrageurs' transaction costs increase after the introduction of the dark venu. This is implied from part 3 of Proposition~\ref{welfare analysis}. 
\end{enumerate}

\subsection{Data} \label{sec data}
We use transaction-level data from  Uniswap and Sushiswap to identify frontrunning arbitrages. We run our own Ethereum node to get access to the blockchain history. A modified geth client is used to export all transaction receipts where a \textit{swap} event was triggered by a smart contract of Uniswap or Sushiswap. Our dataset contains all swap transactions from block number~$10000835$ {created on May 4, 2020} to block number~$12344944$ {created on April 30, 2021}. {For the AMMs transactions in the data,} we follow the method described in ~\cite{Wang2022impact} to identify  frontrunning arbitrages and calculate their revenues. 

We use the API services provided by Flashbots to collect transactions submitted through the private channel to the miners. We collect data starting from February 11, 2021, when the first Flashbots block was mined, till July 31, 2021. This choice eliminates the influence of the new fee mechanism introduced by EIP 1559 after August 2021. 

We acquire the Ethereum block data from Blockchair available at \url{https://gz.blockchair.com/ethereum/blocks/}. The data cover the period from May 1, 2020 to July 31, 2021.  The data include the gas fee revenues earned by miners. 


\subsection{Definition of Variables and Stylized Facts} \label{sec stylized facts}
{We describe the main variables used in our statistical analysis, and provide empirical regularities observed in our data.}

\textbf{{\textit{Dark Venue Adoption Rate of Miners.}}} We estimate the dark venue adoption rate in day $t$ using the  number of blocks mined in day $t$ that contains Flashbots transactions divided by the total number of blocks mined in day $t$. 

\textbf{\textit{Miners'  Revenue per Block.}} If a miner mines a block that contains transactions submitted through Flashbots, then his revenue accounts for Flashbots transactions in this block plus gas fee proceeds from transaction submitted through the mempool. If a miner mines a block that only contains transaction submitted through mempool, then his revenue consists of gas fees paid by those transactions. We do not account for the fixed block reward in our measure of miners' revenue. 

\textbf{\textit{Arbitrageurs' Cost-to-Revenue Ratio .} }For each frontrunning arbitrage order identified, arbitrageur's cost-to-revenue ratio is measured by the gas fee paid by this arbitrageur divided by the revenue of the frontrunning arbitrage. Both the gas fee and arbitrage revenue are in the unit of ether.


\textbf{\textit{Users' Probability of Being Frontrun.}} For each transaction  submitted through the lit venue, we examine whether it is frontrunnable and whether it has been frontrun using a methodology described in Appendix~\ref{sec:fron2}. The probability of being frontrun in day $t$ is  the number of transactions which were frontrun in day $t$ divided by the number of all frontrunnable transactions submitted in that day.

\textbf{\textit{Proportion of Users' Transaction Sent Through the Dark Venue.}} For each transaction submitted through the dark venue, we examine whether it would be frontrunnable if were submitted through the lit venue. The  proportion of transactions sent through the dark venue in day $t$is  the number of frontrunnable transactions submitted through the dark venue during day $t$ divided by the number of all frontrunnable transactions submitted during that day.

\begin{table}[htb]
\begin{adjustbox}{width=\columnwidth,center}
\begin{tabular}{lllllll}
\hline
                                            & N       & Mean  & SD    & 10th  & 50th  & 90th  \\ \hline
\multicolumn{7}{c}{Panel A: Miner Data}                                                       \\ \hline
Daily Dark Venue Adoption Rate              & 171     & 0.343 & 0.239 & 0.01  & 0.346 & 0.613 \\
Revenues of Miners at Dark Venue (ETH)      & 377,366  & 0.972 & 17.82 & 0.235 & 0.606 & 2.2   \\
Proportion of Revenue From Dark Venue (ETH) & 377,366  & 0.139 & 0.148 & 0.024 & 0.086 & 0.326 \\
Revenues of Miners at Lit Venue (ETH)       & 2,582,015 & 1.161 & 9.585 & 0.231 & 0.832 & 2.36  \\ \hline
\multicolumn{7}{c}{Panel B: Arbitrageur Data}                                                 \\ \hline
Arbitrage Revenue in Dark Venue (ETH)       & 29,465   & 0.248 & 0.495 & 0.042 & 0.125 & 0.497 \\
Arbitrage Cost in Dark Venue (ETH)          & 29,465   & 0.182 & 0.363 & 0.032 & 0.092 & 0.371 \\
Cost-to-revenue Ratio of Arbitrageurs in Dark Venue           & 29,465   & 0.755 & 0.151 & 0.51  & 0.801 & 0.901 \\
                                            &         &       &       &       &       &       \\
Arbitrage Revenue in Lit Venue (ETH)        & 394,239  & 0.204 & 0.571 & 0.033 & 0.091 & 0.408 \\
Arbitrage Cost in Lit Venue (ETH)           & 394,239  & 0.04  & 0.093 & 0.004 & 0.023 & 0.069 \\
Cost-to-revenue Ratio of Arbitrageurs in Lit Venue            & 394,239  & 0.309 & 0.239 & 0.021 & 0.261 & 0.662 \\ \hline
\multicolumn{7}{c}{Panel C: User Data}                                                        \\ \hline
Daily Probability of Being Attacked         & 80      & 0.165 & 0.034 & 0.120 & 0.165 & 0.209 \\
Daily Ratio of Using Dark Venue             & 80      & 0.033 & 0.038 & 0     & 0.01  & 0.09  \\ \hline
\end{tabular}
\end{adjustbox}
\caption{Summary statistics of the data set}\label{tab:em}
\end{table}

\begin{figure}[htbp!]
    \centering
    \begin{minipage}{.5\textwidth}
\centering
    \begin{tikzpicture}[scale  = 0.8]
\begin{axis}[clip mode=individual, date coordinates in=x,
    xticklabel=\year-\month, xlabel={Date (YYYY-MM)},     ylabel={Dark Venue Adoption Rate of Miners}, ymin=0, ymax=0.8, 
x label style={
    at={(0.5,-.1)},
    anchor=south,
}, y label style={
    at={(.03,0.0)},
    anchor=west,
}, title style={at={(0.5,1.06)},anchor=north,},
]
\addplot+[black] coordinates { (2021-2-11,0.006445) (2021-2-12,0.006131) (2021-2-13,0.005605) (2021-2-14,0.007776) (2021-2-15,0.005080) (2021-2-16,0.006939) (2021-2-17,0.011758) (2021-2-18,0.010237) (2021-2-19,0.008819) (2021-2-20,0.011085) (2021-2-21,0.010023) (2021-2-22,0.008159) (2021-2-23,0.006767) (2021-2-24,0.010357) (2021-2-25,0.010284) (2021-2-26,0.010547) (2021-2-27,0.010633) (2021-2-28,0.011133) (2021-3-01,0.008338) (2021-3-02,0.006412) (2021-3-03,0.004526) (2021-3-04,0.003223) (2021-3-05,0.002605) (2021-3-06,0.003962) (2021-3-07,0.006176) (2021-3-08,0.006154) (2021-3-09,0.009949) (2021-3-10,0.013472) (2021-3-11,0.014186) (2021-3-12,0.016383) (2021-3-13,0.015356) (2021-3-14,0.017633) (2021-3-15,0.022837) (2021-3-16,0.024847) (2021-3-17,0.021437) (2021-3-18,0.021228) (2021-3-19,0.026283) (2021-3-20,0.028957) (2021-3-21,0.040959) (2021-3-22,0.039753) (2021-3-23,0.049931) (2021-3-24,0.063078) (2021-3-25,0.053423) (2021-3-26,0.045413) (2021-3-27,0.045181) (2021-3-28,0.030802) (2021-3-29,0.033943) (2021-3-30,0.048474) (2021-3-31,0.107099) (2021-4-01,0.134592) (2021-4-02,0.147350) (2021-4-03,0.133905) (2021-4-04,0.213535) (2021-4-05,0.275115) (2021-4-06,0.312818) (2021-4-07,0.325685) (2021-4-08,0.323160) (2021-4-09,0.325538) (2021-4-10,0.293224) (2021-4-11,0.255881) (2021-4-12,0.297268) (2021-4-13,0.304294) (2021-4-14,0.323151) (2021-4-15,0.233385) (2021-4-16,0.248096) (2021-4-17,0.231019) (2021-4-18,0.195508) (2021-4-19,0.220938) (2021-4-20,0.281763) (2021-4-21,0.304034) (2021-4-22,0.331463) (2021-4-23,0.365426) (2021-4-24,0.327385) (2021-4-25,0.296513) (2021-4-26,0.346314) (2021-4-27,0.326603) (2021-4-28,0.327866) (2021-4-29,0.343615) (2021-4-30,0.331887) (2021-5-01,0.345200) (2021-5-02,0.349797) (2021-5-03,0.338405) (2021-5-04,0.352706) (2021-5-05,0.321724) (2021-5-06,0.296783) (2021-5-07,0.283853) (2021-5-08,0.302035) (2021-5-09,0.324637) (2021-5-10,0.359368) (2021-5-11,0.331162) (2021-5-12,0.360770) (2021-5-13,0.352554) (2021-5-14,0.390459) (2021-5-15,0.431109) (2021-5-16,0.404769) (2021-5-17,0.449291) (2021-5-18,0.437413) (2021-5-19,0.401837) (2021-5-20,0.454433) (2021-5-21,0.481279) (2021-5-22,0.484061) (2021-5-23,0.458320) (2021-5-24,0.495144) (2021-5-25,0.467989) (2021-5-26,0.521381) (2021-5-27,0.527938) (2021-5-28,0.550983) (2021-5-29,0.520953) (2021-5-30,0.534518) (2021-5-31,0.539000) (2021-6-01,0.535505) (2021-6-02,0.587765) (2021-6-03,0.628826) (2021-6-04,0.639055) (2021-6-05,0.631800) (2021-6-06,0.602419) (2021-6-07,0.651998) (2021-6-08,0.606023) (2021-6-09,0.598415) (2021-6-10,0.606159) (2021-6-11,0.628447) (2021-6-12,0.614859) (2021-6-13,0.577939) (2021-6-14,0.580314) (2021-6-15,0.559877) (2021-6-16,0.584935) (2021-6-17,0.548362) (2021-6-18,0.540827) (2021-6-19,0.577653) (2021-6-20,0.545525) (2021-6-21,0.573306) (2021-6-22,0.490135) (2021-6-23,0.619536) (2021-6-24,0.606887) (2021-6-25,0.570640) (2021-6-26,0.600747) (2021-6-27,0.562810) (2021-6-28,0.570987) (2021-6-29,0.587730) (2021-6-30,0.581100) (2021-7-01,0.603150) (2021-7-02,0.599192) (2021-7-03,0.624591) (2021-7-04,0.612264) (2021-7-05,0.619723) (2021-7-06,0.577121) (2021-7-07,0.569327) (2021-7-08,0.569922) (2021-7-09,0.586169) (2021-7-10,0.598728) (2021-7-11,0.604858) (2021-7-12,0.554022) (2021-7-13,0.575234) (2021-7-14,0.612676) (2021-7-15,0.603392) (2021-7-16,0.567282) (2021-7-17,0.599654) (2021-7-18,0.585465) (2021-7-19,0.595338) (2021-7-20,0.605301) (2021-7-21,0.645463) (2021-7-22,0.577612) (2021-7-23,0.638670) (2021-7-24,0.668267) (2021-7-25,0.672395) (2021-7-26,0.644683) (2021-7-27,0.618448) (2021-7-28,0.625974) (2021-7-29,0.647892) (2021-7-30,0.609861) (2021-7-31,0.593681)

  };
\end{axis}

\end{tikzpicture}
    \setcaptionwidth{0.9\linewidth}
    \caption{Adoption rate of Flashbots.}
    \label{fig:miner1}
    \end{minipage}%
    \begin{minipage}{0.5\textwidth}
      \centering
    \begin{tikzpicture}[scale  = 0.8]
\begin{axis}[clip mode=individual, date coordinates in=x,
    xticklabel=\year-\month, xlabel={Date ((YYYY-MM))},     ylabel={Proportion of Revenue from Dark}, ymin=0.00, ymax=0.23, 
x label style={
    at={(0.5,-.1)},
    anchor=south,
},  yticklabel style={
        /pgf/number format/fixed,
        /pgf/number format/precision=5
},
scaled y ticks=false,
title style={at={(0.5,1.06)},anchor=north,},
]
\addplot[scatter,opacity=0.5] coordinates { (2021-02-11,0.023576)(2021-2-12,0.014914)(2021-2-13,0.016545)(2021-2-14,0.012258)(2021-2-15,0.025719)(2021-2-16,0.011607)(2021-2-17,0.033571)(2021-2-18,0.026463)(2021-2-19,0.029088)(2021-2-20,0.036338)(2021-2-21,0.033429)(2021-2-22,0.034206)(2021-2-23,0.041914)(2021-2-24,0.043112)(2021-2-25,0.040652)(2021-2-26,0.040832)(2021-2-27,0.0226)(2021-2-28,0.031697)(2021-3-1,0.048843)(2021-3-2,0.051984)(2021-3-3,0.022891)(2021-3-4,0.038354)(2021-3-5,0.081062)(2021-3-6,0.065751)(2021-3-7,0.050454)(2021-3-8,0.092142)(2021-3-9,0.070399)(2021-3-10,0.078622)(2021-3-11,0.072256)(2021-3-12,0.072743)(2021-3-13,0.053032)(2021-3-14,0.08214)(2021-3-15,0.067313)(2021-3-16,0.074781)(2021-3-17,0.070651)(2021-3-18,0.064809)(2021-3-19,0.07687)(2021-3-20,0.076139)(2021-3-21,0.091947)(2021-3-22,0.086495)(2021-3-23,0.087649)(2021-3-24,0.091275)(2021-3-25,0.094633)(2021-3-26,0.093581)(2021-3-27,0.10136)(2021-3-28,0.105843)(2021-3-29,0.100303)(2021-3-30,0.08383)(2021-3-31,0.088104)(2021-4-1,0.090025)(2021-4-2,0.083494)(2021-4-3,0.084205)(2021-4-4,0.09832)(2021-4-5,0.083704)(2021-4-6,0.081613)(2021-4-7,0.085354)(2021-4-8,0.09401)(2021-4-9,0.101578)(2021-4-10,0.10528)(2021-4-11,0.099457)(2021-4-12,0.094289)(2021-4-13,0.093468)(2021-4-14,0.083388)(2021-4-15,0.098367)(2021-4-16,0.082477)(2021-4-17,0.099413)(2021-4-18,0.087439)(2021-4-19,0.084252)(2021-4-20,0.08276)(2021-4-21,0.088447)(2021-4-22,0.09802)(2021-4-23,0.112521)(2021-4-24,0.113894)(2021-4-25,0.126747)(2021-4-26,0.120713)(2021-4-27,0.1198)(2021-4-28,0.120696)(2021-4-29,0.108107)(2021-4-30,0.130203)(2021-5-1,0.139014)(2021-5-2,0.197992)(2021-5-3,0.138836)(2021-5-4,0.113262)(2021-5-5,0.117765)(2021-5-6,0.092625)(2021-5-7,0.093128)(2021-5-8,0.095052)(2021-5-9,0.082003)(2021-5-10,0.081798)(2021-5-11,0.062317)(2021-5-12,0.073362)(2021-5-13,0.080286)(2021-5-14,0.08487)(2021-5-15,0.099928)(2021-5-16,0.117801)(2021-5-17,0.108421)(2021-5-18,0.106013)(2021-5-19,0.098359)(2021-5-20,0.104055)(2021-5-21,0.108471)(2021-5-22,0.119586)(2021-5-23,0.120761)(2021-5-24,0.118564)(2021-5-25,0.128093)(2021-5-26,0.159287)(2021-5-27,0.162245)(2021-5-28,0.163018)(2021-5-29,0.152343)(2021-5-30,0.150779)(2021-5-31,0.150487)(2021-6-1,0.140471)(2021-6-2,0.167172)(2021-6-3,0.170587)(2021-6-4,0.189273)(2021-6-5,0.161935)(2021-6-6,0.158583)(2021-6-7,0.164068)(2021-6-8,0.162787)(2021-6-9,0.17561)(2021-6-10,0.181029)(2021-6-11,0.181088)(2021-6-12,0.180342)(2021-6-13,0.172289)(2021-6-14,0.162977)(2021-6-15,0.157784)(2021-6-16,0.163054)(2021-6-17,0.166058)(2021-6-18,0.166954)(2021-6-19,0.163052)(2021-6-20,0.147643)(2021-6-21,0.146332)(2021-6-22,0.136335)(2021-6-23,0.162467)(2021-6-24,0.15416)(2021-6-25,0.159661)(2021-6-26,0.173536)(2021-6-27,0.164834)(2021-6-28,0.159369)(2021-6-29,0.151828)(2021-6-30,0.148917)(2021-7-1,0.160607)(2021-7-2,0.167671)(2021-7-3,0.193077)(2021-7-4,0.164797)(2021-7-5,0.164607)(2021-7-6,0.140493)(2021-7-7,0.120136)(2021-7-8,0.132436)(2021-7-9,0.122689)(2021-7-10,0.16376)(2021-7-11,0.152754)(2021-7-12,0.138147)(2021-7-13,0.145095)(2021-7-14,0.151127)(2021-7-15,0.156267)(2021-7-16,0.131504)(2021-7-17,0.138036)(2021-7-18,0.138718)(2021-7-19,0.136375)(2021-7-20,0.155874)(2021-7-21,0.150756)(2021-7-22,0.135477)(2021-7-23,0.148578)(2021-7-24,0.170588)(2021-7-25,0.158782)(2021-7-26,0.155392)(2021-7-27,0.143923)(2021-7-28,0.143062)(2021-7-29,0.147015)(2021-7-30,0.142925)(2021-7-31,0.146985) };

\end{axis}

\end{tikzpicture}
    \setcaptionwidth{0.9\linewidth}
    \caption{Proportion of Flashbots miners' revenue from dark venue.}
    \label{fig:miner3}
    \end{minipage}
\end{figure}

\begin{figure}[htbp!]
\centering
\begin{subfigure}{.49\textwidth}
  \centering
  \includegraphics[width= \linewidth]{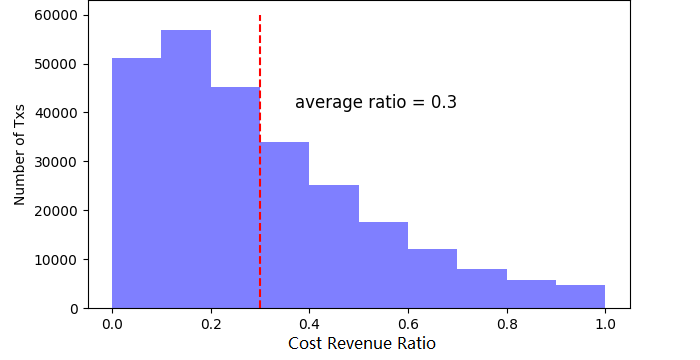}
  \label{fig:lit_dis_before}
\end{subfigure}%
\begin{subfigure}{.49\textwidth}
  \centering
  \includegraphics[width = \linewidth]{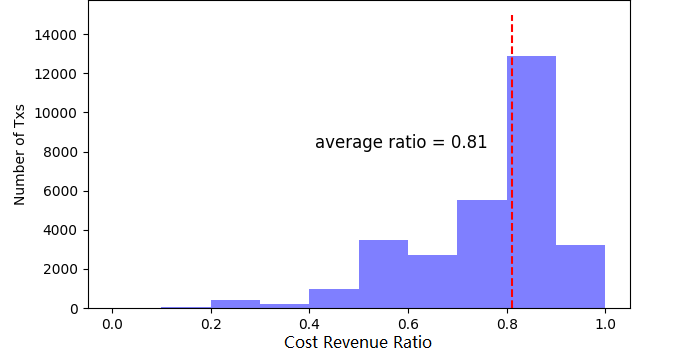}
  \label{fig:dark_dis_att} 
\end{subfigure}
\begin{subfigure}{.49\textwidth}
  \centering
  \includegraphics[width = \linewidth]{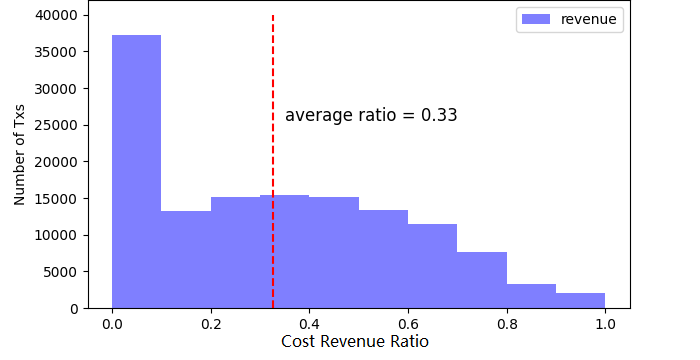}
   \label{fig:lit_dis_after} 
\end{subfigure}
\caption{Panel (a) - top left: Distribution of the cost-to-revenue ratio of attackers in the lit venue before the introduction of the dark venue. Panel (b) - top right: Distribution of the cost-to-revenue ratio of attackers in the dark venue. Panel (c) - bottom: Distribution of the cost-to-revenue ratio of attackers in the lit venue after the introduction of the dark venue.
}
\label{fig:arbitrageurs}
\end{figure}
\textbf{\textit{Descriptive Statistics and Stylized Facts.}} Table~\ref{tab:em}  presents summary statistics of the data. Figure~\ref{fig:miner1} plots the estimated adoption rate of dark venue. For miners who join the dark venue, we plot the proportion of extracted revenue in Figure~\ref{fig:miner3}. We can clearly observe that dark venue transactions contribute a nontrivial (around 15\%) portion to the revenues of miners who joined dark venue. The distribution of cost-to-revenue ratio of arbitrageurs {is} plotted in Figure~\ref{fig:arbitrageurs}. Comparing panel (a)-(c), we observe that the cost-to-revenue ratio for arbitrageurs who submit through the dark venue is skewed right and higher than that of arbitrageur who use lit venue. The average cost-to-revenue ratio increases after the introduction of the dark venue. Figure~\ref{fig:attacker2} plots the daily average cost-to-revenue ratio of arbitrageurs in the lit and the dark venue. After the introduction of the dark venue, the cost-to-revenue ratio in the dark venue steadily increases while the cost-to-revenue ratio in the lit venue decreases. Our model offers a  plausible explanation to this observed pattern: as the miner adoption rate of the dark venue increases, more arbitrageurs migrate from the lit venue to the dark venue, which increases competition and raises transaction costs. Recall that transactions sent through the dark venue face execution risk. When the block is not mined by miners who join the dark venue, arbitrage transactions sent through the lit venue are executed, and the transaction cost is lower because of the smaller competition. Figure~\ref{fig:user} plots users' probability of being frontrun (red) and proportion of users' transaction submitted to the dark venue (black). The graph suggests that users may migrate to the dark venue when the frontrunning risk they face increases.

\begin{figure}[!htb]
    \centering
    \begin{minipage}{.5\textwidth}
   \setcaptionwidth{0.9\linewidth}
    \begin{tikzpicture}[scale  = 0.8]
\begin{axis}[clip mode=individual, date coordinates in=x,
    xticklabel=\year-\month, xlabel={Date (YYYY-MM)},     ylabel={Daily Average Cost Revenue Ratio}, ymin=0, ymax=1, 
x label style={
    at={(0.5,-.1)},
    anchor=south,
}, y label style={
    at={(.03,0.0)},
    anchor=west,
}, title style={at={(0.5,1.06)},anchor=north,},
]
\addplot[mark=none, black, line width=1pt] coordinates { (2021-03-06,0.342000)(2021-03-07,0.403375)(2021-03-08,0.616668)(2021-03-09,0.568181)(2021-03-10,0.555172)(2021-03-11,0.542857)(2021-03-12,0.615667)(2021-03-13,0.621212)(2021-03-14,0.592460)(2021-03-15,0.594315)(2021-03-16,0.633748)(2021-03-17,0.628416)(2021-03-18,0.634599)(2021-03-19,0.663100)(2021-03-20,0.663527)(2021-03-21,0.664077)(2021-03-22,0.699202)(2021-03-23,0.699641)(2021-03-24,0.712581)(2021-03-25,0.742722)(2021-03-26,0.686411)(2021-03-27,0.676105)(2021-03-28,0.732993)(2021-03-29,0.741827)(2021-03-30,0.692954)(2021-03-31,0.712844)(2021-04-01,0.660109)(2021-04-02,0.670801)(2021-04-03,0.643648)(2021-04-04,0.633331)(2021-04-05,0.686601)(2021-04-06,0.692733)(2021-04-07,0.691188)(2021-04-08,2.053049)(2021-04-09,0.769808)(2021-04-10,0.752215)(2021-04-11,0.730011)(2021-04-12,0.713897)(2021-04-13,0.720730)(2021-04-14,0.748133)(2021-04-15,0.746414)(2021-04-16,0.732949)(2021-04-17,0.772469)(2021-04-18,0.775735)(2021-04-19,0.764068)(2021-04-20,0.769164)(2021-04-21,0.796488)(2021-04-22,0.812664)(2021-04-23,0.842074)(2021-04-24,0.858507)(2021-04-25,0.883807)(2021-04-26,0.864172)(2021-04-27,0.846912)(2021-04-28,0.849458)(2021-04-29,0.856939)(2021-04-30,0.933476)(2021-05-01,0.844499) };
\addplot[mark=none, blue, line width=1pt] coordinates { (2020-05-25,0.277095)(2020-05-30,0.543071)(2020-05-31,0.253867)(2020-06-01,0.406021)(2020-06-02,0.211277)(2020-06-03,0.287571)(2020-06-04,0.264877)(2020-06-05,0.268867)(2020-06-06,0.241797)(2020-06-07,0.283194)(2020-06-08,0.295761)(2020-06-09,0.283527)(2020-06-10,0.325766)(2020-06-11,0.337569)(2020-06-12,0.305279)(2020-06-13,0.289616)(2020-06-14,0.285435)(2020-06-15,0.344953)(2020-06-16,0.327831)(2020-06-17,0.315333)(2020-06-18,0.308862)(2020-06-19,0.337376)(2020-06-20,0.304903)(2020-06-21,0.328822)(2020-06-22,0.336836)(2020-06-23,0.297847)(2020-06-24,0.287304)(2020-06-25,0.310968)(2020-06-26,0.323071)(2020-06-27,0.298156)(2020-06-28,0.278302)(2020-06-29,0.289674)(2020-06-30,0.323956)(2020-07-01,0.309949)(2020-07-02,0.271396)(2020-07-03,0.296538)(2020-07-04,0.246732)(2020-07-05,0.241356)(2020-07-06,0.263243)(2020-07-07,0.262557)(2020-07-08,0.312931)(2020-07-09,0.288914)(2020-07-10,0.285849)(2020-07-11,0.298108)(2020-07-12,0.279451)(2020-07-13,0.279100)(2020-07-14,0.273326)(2020-07-15,0.271079)(2020-07-16,0.303955)(2020-07-17,0.292699)(2020-07-18,0.284222)(2020-07-19,0.300088)(2020-07-20,0.296583)(2020-07-21,0.310587)(2020-07-22,0.343623)(2020-07-23,0.354033)(2020-07-24,0.372898)(2020-07-25,0.369569)(2020-07-26,0.367675)(2020-07-27,0.400708)(2020-07-28,0.317201)(2020-07-29,0.298841)(2020-07-30,0.292337)(2020-07-31,0.276699)(2020-08-01,0.294887)(2020-08-02,0.360283)(2020-08-03,0.312674)(2020-08-04,0.283161)(2020-08-05,0.257921)(2020-08-06,0.283900)(2020-08-07,0.282052)(2020-08-08,0.282718)(2020-08-09,0.277173)(2020-08-10,0.286321)(2020-08-11,0.289626)(2020-08-12,0.318800)(2020-08-13,0.309910)(2020-08-14,0.257054)(2020-08-15,0.241042)(2020-08-16,0.216760)(2020-08-17,0.211332)(2020-08-18,0.199727)(2020-08-19,0.228650)(2020-08-20,0.226086)(2020-08-21,0.239281)(2020-08-22,0.198270)(2020-08-23,0.193564)(2020-08-24,0.184406)(2020-08-25,0.218115)(2020-08-26,0.237839)(2020-08-27,0.242804)(2020-08-28,0.241157)(2020-08-29,0.243703)(2020-08-30,0.323839)(2020-08-31,0.310871)(2020-09-01,0.302393)(2020-09-02,0.218875)(2020-09-03,0.221396)(2020-09-04,0.224192)(2020-09-05,0.230601)(2020-09-06,0.206103)(2020-09-07,0.187201)(2020-09-08,0.202811)(2020-09-09,0.224146)(2020-09-10,0.272281)(2020-09-11,0.244671)(2020-09-12,0.211224)(2020-09-13,0.219771)(2020-09-14,0.252496)(2020-09-15,0.260896)(2020-09-16,0.248740)(2020-09-17,0.285134)(2020-09-18,0.328287)(2020-09-19,0.251626)(2020-09-20,0.221951)(2020-09-21,0.246140)(2020-09-22,0.268908)(2020-09-23,0.239471)(2020-09-24,0.224317)(2020-09-25,0.239325)(2020-09-26,0.232483)(2020-09-27,0.253897)(2020-09-28,0.231457)(2020-09-29,0.245467)(2020-09-30,0.229227)(2020-10-01,0.221601)(2020-10-02,0.220813)(2020-10-03,0.238198)(2020-10-04,0.241759)(2020-10-05,0.266324)(2020-10-06,0.275422)(2020-10-07,0.269584)(2020-10-08,0.267707)(2020-10-09,0.256491)(2020-10-10,0.256478)(2020-10-11,0.259529)(2020-10-12,0.268531)(2020-10-13,0.273610)(2020-10-14,0.277586)(2020-10-15,0.267557)(2020-10-16,0.314927)(2020-10-17,0.262868)(2020-10-18,0.230604)(2020-10-19,0.237213)(2020-10-20,0.280338)(2020-10-21,0.298370)(2020-10-22,0.290063)(2020-10-23,0.273640)(2020-10-24,0.225352)(2020-10-25,0.236383)(2020-10-26,0.285029)(2020-10-27,0.270133)(2020-10-28,0.264196)(2020-10-29,0.276209)(2020-10-30,0.272644)(2020-10-31,0.232733)(2020-11-01,0.228251)(2020-11-02,0.291039)(2020-11-03,0.258473)(2020-11-04,0.257871)(2020-11-05,0.290437)(2020-11-06,0.269973)(2020-11-07,0.276109)(2020-11-08,0.219699)(2020-11-09,0.238190)(2020-11-10,0.257404)(2020-11-11,0.302264)(2020-11-12,0.322301)(2020-11-13,0.277327)(2020-11-14,0.255745)(2020-11-15,0.229210)(2020-11-16,0.268345)(2020-11-17,0.301878)(2020-11-18,0.319851)(2020-11-19,0.255104)(2020-11-20,0.294310)(2020-11-21,0.291821)(2020-11-22,0.283180)(2020-11-23,0.284004)(2020-11-24,0.300614)(2020-11-25,0.297122)(2020-11-26,0.307532)(2020-11-27,0.252682)(2020-11-28,0.261390)(2020-11-29,0.277329)(2020-11-30,0.311874)(2020-12-01,0.314746)(2020-12-02,0.316974)(2020-12-03,0.295733)(2020-12-04,0.317112)(2020-12-05,0.271380)(2020-12-06,0.258622)(2020-12-07,0.286992)(2020-12-08,0.318001)(2020-12-09,0.338741)(2020-12-10,0.323225)(2020-12-11,0.316605)(2020-12-12,0.313302)(2020-12-13,0.319542)(2020-12-14,0.354502)(2020-12-15,0.383533)(2020-12-16,0.379739)(2020-12-17,0.402253)(2020-12-18,0.376960)(2020-12-19,0.316954)(2020-12-20,0.283122)(2020-12-21,0.341419)(2020-12-22,0.345246)(2020-12-23,0.401195)(2020-12-24,0.384556)(2020-12-25,0.388788)(2020-12-26,0.377930)(2020-12-27,0.396848)(2020-12-28,0.398112)(2020-12-29,0.421913)(2020-12-30,0.442935)(2020-12-31,0.404061)(2021-01-01,0.363167)(2021-01-02,0.392272)(2021-01-03,0.429174)(2021-01-04,0.437287)(2021-01-05,0.423954)(2021-01-06,0.428226)(2021-01-07,0.430033)(2021-01-08,0.438059)(2021-01-09,0.422379)(2021-01-10,0.387048)(2021-01-11,0.456748)(2021-01-12,0.404620)(2021-01-13,0.372586)(2021-01-14,0.385004)(2021-01-15,0.378747)(2021-01-16,0.379010)(2021-01-17,0.353058)(2021-01-18,0.379203)(2021-01-19,0.400815)(2021-01-20,0.344222)(2021-01-21,0.346281)(2021-01-22,0.365162)(2021-01-23,0.348722)(2021-01-24,0.339759)(2021-01-25,0.348620)(2021-01-26,0.366997)(2021-01-27,0.370811)(2021-01-28,0.334556)(2021-01-29,0.364820)(2021-01-30,0.339005)(2021-01-31,0.372690)(2021-02-01,0.407186)(2021-02-02,0.437855)(2021-02-03,0.429981)(2021-02-04,0.457152)(2021-02-05,0.451047)(2021-02-06,0.466585)(2021-02-07,0.440221)(2021-02-08,0.435634)(2021-02-09,0.450931)(2021-02-10,0.440791)(2021-02-11,0.425224) };
\addplot[mark=none, orange, line width=1pt] coordinates { (2021-02-11,0.446901)(2021-02-12,0.429959)(2021-02-13,0.455433)(2021-02-14,0.419697)(2021-02-15,0.412785)(2021-02-16,0.420747)(2021-02-17,0.420318)(2021-02-18,0.416601)(2021-02-19,0.414921)(2021-02-20,0.415482)(2021-02-21,0.403983)(2021-02-22,0.404967)(2021-02-23,0.426133)(2021-02-24,0.410554)(2021-02-25,0.402705)(2021-02-26,0.389832)(2021-02-27,0.353609)(2021-02-28,0.361870)(2021-03-01,0.374160)(2021-03-02,0.384527)(2021-03-03,0.366731)(2021-03-04,0.383343)(2021-03-05,0.385434)(2021-03-06,0.360587)(2021-03-07,0.376503)(2021-03-08,0.401053)(2021-03-09,0.405772)(2021-03-10,0.412968)(2021-03-11,0.408748)(2021-03-12,0.411516)(2021-03-13,0.374312)(2021-03-14,0.392751)(2021-03-15,0.383139)(2021-03-16,0.341911)(2021-03-17,0.336369)(2021-03-18,0.317715)(2021-03-19,0.308987)(2021-03-20,0.292789)(2021-03-21,0.306878)(2021-03-22,0.292035)(2021-03-23,0.313084)(2021-03-24,0.316600)(2021-03-25,0.323095)(2021-03-26,0.280667)(2021-03-27,0.269861)(2021-03-28,0.301655)(2021-03-29,0.281821)(2021-03-30,0.298105)(2021-03-31,0.300279)(2021-04-01,0.322714)(2021-04-02,0.343460)(2021-04-03,0.298807)(2021-04-04,0.271173)(2021-04-05,0.246399)(2021-04-06,0.238363)(2021-04-07,0.257951)(2021-04-08,0.218766)(2021-04-09,0.223755)(2021-04-10,0.214012)(2021-04-11,0.175341)(2021-04-12,0.173990)(2021-04-13,0.194847)(2021-04-14,0.207834)(2021-04-15,0.208716)(2021-04-16,0.272202)(2021-04-17,0.233739)(2021-04-18,0.277257)(2021-04-19,0.289878)(2021-04-20,0.298859)(2021-04-21,0.306528)(2021-04-22,0.283168)(2021-04-23,0.231783)(2021-04-24,0.190118)(2021-04-25,0.152387)(2021-04-26,0.149638)(2021-04-27,0.152730)(2021-04-28,0.196422)(2021-04-29,0.173576)(2021-04-30,0.140932)(2021-05-01,0.112775) };

\end{axis}

\end{tikzpicture}
    \caption{Daily average cost-to-revenue ratio of arbitrageurs. Blue: arbitrageurs in lit venue before the introduction of the dark venue, Black: arbitrageurs in dark venue, Orange: arbitrageurs in lit venue after the introduction of the dark venue.}
    \label{fig:attacker2}
    \end{minipage}%
    \begin{minipage}{0.5\textwidth}
      \centering \setcaptionwidth{0.9\linewidth}
 \begin{tikzpicture}[scale  = 0.8]
\begin{axis}[clip mode=individual, date coordinates in=x, axis y line*=left,
    xticklabel=\year-\month, xlabel={Date (YYYY-MM)},
    log basis y={10},ylabel={Probability of Being Frontrun}, ymin=0, ymax=0.3, 
x label style={
    at={(0.5,-.1)},
    anchor=south,
}, y label style={
    at={(.03,0.05)},
    anchor=west,
}, title style={at={(0.5,1.06)},anchor=north,},
]
\addplot[mark=none, black, line width=1pt] coordinates { (2021-05-01,0.231820)(2021-04-30,0.174137)(2021-04-29,0.174243)(2021-04-28,0.185247)(2021-04-27,0.185799)(2021-04-26,0.173952)(2021-04-25,0.148847)(2021-04-24,0.183684)(2021-04-23,0.182953)(2021-04-22,0.156447)(2021-04-21,0.108881)(2021-04-20,0.109619)(2021-04-19,0.142068)(2021-04-18,0.161679)(2021-04-17,0.167704)(2021-04-16,0.183144)(2021-04-15,0.249591)(2021-04-14,0.207576)(2021-04-13,0.212796)(2021-04-12,0.220902)(2021-04-11,0.234626)(2021-04-10,0.207658)(2021-04-09,0.221772)(2021-04-08,0.200672)(2021-04-07,0.166988)(2021-04-06,0.180467)(2021-04-05,0.190212)(2021-04-04,0.212053)(2021-04-03,0.168114)(2021-04-02,0.158203)(2021-04-01,0.166760)(2021-03-31,0.157291)(2021-03-30,0.177489)(2021-03-29,0.184603)(2021-03-28,0.208510)(2021-03-27,0.196841)(2021-03-26,0.175856)(2021-03-25,0.166667)(2021-03-24,0.156461)(2021-03-23,0.160476)(2021-03-22,0.155092)(2021-03-21,0.137146)(2021-03-20,0.143798)(2021-03-19,0.142001)(2021-03-18,0.126193)(2021-03-17,0.118777)(2021-03-16,0.123256)(2021-03-15,0.113134)(2021-03-14,0.135009)(2021-03-13,0.120340)(2021-03-12,0.110465)(2021-03-11,0.152297)(2021-03-10,0.150038)(2021-03-09,0.142721)(2021-03-08,0.149746)(2021-03-07,0.158643)(2021-03-06,0.203021)(2021-03-05,0.185562)(2021-03-04,0.178806)(2021-03-03,0.182417)(2021-03-02,0.172345)(2021-03-01,0.200000)(2021-02-28,0.215530)(2021-02-27,0.202708)(2021-02-26,0.189768)(2021-02-25,0.192613)(2021-02-24,0.138210)(2021-02-23,0.117466)(2021-02-22,0.106844)(2021-02-21,0.135432)(2021-02-20,0.129589)(2021-02-19,0.124894)(2021-02-18,0.120509)(2021-02-17,0.130140)(2021-02-16,0.140814)(2021-02-15,0.144981)(2021-02-14,0.141540)(2021-02-13,0.115978)(2021-02-12,0.125641)(2021-02-11,0.163982) };
\end{axis}

\begin{axis}[clip mode=individual, date coordinates in=x, axis x line = none,
     axis y line*=right,ylabel=\color{red}{Propotion of Using Dark Venue}, ymin=0, ymax=0.3, 
 y label style={
    at={(1.3,0.05)},
    anchor=west,
}
]
\addplot[mark=none, red, line width=1pt] coordinates { (2021-05-01,0.102088)(2021-04-30,0.079097)(2021-04-29,0.085007)(2021-04-28,0.090468)(2021-04-27,0.085003)(2021-04-26,0.081611)(2021-04-25,0.067485)(2021-04-24,0.089073)(2021-04-23,0.089621)(2021-04-22,0.071243)(2021-04-21,0.055157)(2021-04-20,0.055713)(2021-04-19,0.052150)(2021-04-18,0.054282)(2021-04-17,0.073057)(2021-04-16,0.076146)(2021-04-15,0.086460)(2021-04-14,0.102252)(2021-04-13,0.101531)(2021-04-12,0.104622)(2021-04-11,0.099662)(2021-04-10,0.091911)(2021-04-09,0.101456)(2021-04-08,0.087777)(2021-04-07,0.078979)(2021-04-06,0.081758)(2021-04-05,0.075585)(2021-04-04,0.065816)(2021-04-03,0.042214)(2021-04-02,0.044493)(2021-04-01,0.043264)(2021-03-31,0.027627)(2021-03-30,0.013469)(2021-03-29,0.009343)(2021-03-28,0.009352)(2021-03-27,0.013715)(2021-03-26,0.014062)(2021-03-25,0.015162)(2021-03-24,0.017602)(2021-03-23,0.016736)(2021-03-22,0.012612)(2021-03-21,0.011041)(2021-03-20,0.008201)(2021-03-19,0.005956)(2021-03-18,0.005091)(2021-03-17,0.004301)(2021-03-16,0.008057)(2021-03-15,0.007407)(2021-03-14,0.005977)(2021-03-13,0.005961)(2021-03-12,0.007449)(2021-03-11,0.005721)(2021-03-10,0.005026)(2021-03-09,0.003099)(2021-03-08,0.001224)(2021-03-07,0.001411)(2021-03-06,0.000364)(2021-03-05,0.000155)(2021-03-04,0.000000)(2021-03-03,0.000000)(2021-03-02,0.000000)(2021-03-01,0.000457)(2021-02-28,0.000252)(2021-02-27,0.000000)(2021-02-26,0.000468)(2021-02-25,0.000635)(2021-02-24,0.000000)(2021-02-23,0.000586)(2021-02-22,0.000385)(2021-02-21,0.000000)(2021-02-20,0.000000)(2021-02-19,0.000188)(2021-02-18,0.000267)(2021-02-17,0.000158)(2021-02-16,0.000000)(2021-02-15,0.000000)(2021-02-14,0.000000)(2021-02-13,0.000000)(2021-02-12,0.000000)(2021-02-11,0.000000) };
\end{axis}

\end{tikzpicture}
     \caption{The black line represents the daily average probability of being attacked for frontrunnable users. The red line represents the daily proportion of frontrunnable transactions sent to dark venue.}
     \label{fig:user}
    \end{minipage}
\end{figure}

\subsection{Empirical Results} \label{sec empirical results list}

\subsubsection{The Adoption of Dark Venue.}
The average adoption rate from February to March is 0.02, with a standard deviation of 0.02.
The average adoption rate from April to May is 0.348, with a standard deviation of 0.01.
The average adoption rate from June to July is 0.597, with a standard deviation of 0.033. These estimates are supportive of our model prediction that the dark venue will be at least partially adopted.

\subsubsection{Revenue of Miners in the Dark Venue and in the Lit Venue.}
We estimate the following linear model to compare revenues of miners who adopt the dark venue with revenues of miners who stay in the lit venue:
\begin{equation}\label{regression1}
    MinerRevenue_{t} =  \gamma_t + \rho_1 \ind_{Dark} + \epsilon_{t},
\end{equation}
where $t$ indexes the date, $MinerRevenue_{t}$ is the revenue of miner per block, $\gamma_t$ is the day fixed effects, $\ind_{Dark}$ is a dummy variable for Flashbots blocks, and $\epsilon_{t}$ is an error term. We cluster our standard errors at the day level. The coefficient $\rho_1$ quantifies the sensitivity of  miner's revenue per block on whether he joins the dark venue. 

Table~\ref{regression: minerrev} indicates that 
joining the dark venue on average increases miners' revenue by around 0.16 ETH per block. 
 This is supportive of our model prediction {that the expected payoff of miner who join the dark venue is higher than the expected payoff of miners who stay in the lit venue.} {In addition, the coefficient estimates reveal that these relationships are statistically and  economically significant.}

\begin{table}[t!] \centering
\caption{
 Results from regressing a binary variable, indicating whether or not the miner of the block joins the dark venue, on miner's revenue from mining the block. The regression data covers a sample period from Nov 1, 2020 to Jul 31, 2021.  
 Time fixed effects  are included for all regressions. Standard errors are clustered at the day level. Asterisks denote significance levels (***=1\%, **=5\%, *=10\%).} \label{regression: minerrev}

\begin{tabular*}{0.8\textwidth}{@{\extracolsep{\fill}}  c  c}
\\[-1.8ex]\hline
\hline \\[-1.8ex]
& {\textit{Dependent variables:  Miner's Revenue per Block}} \\
\hline \\[-1.8ex]
 Intercept & 1.21$^{***}$ \\
  & (0.06)  \\
  Dark & 0.16$^{***}$  \\
  & (0.032)  \\ 
\hline \\[-1.8ex]
Day fixed effects? & yes   \\
 Observations & 1,762,017	 \\
 $R^2$ & 0.02  \\
\hline
\hline \\[-1.8ex]
\textit{Note:} & {$^{*}$p$<$0.1; $^{**}$p$<$0.05; $^{***}$p$<$0.01} \\
\end{tabular*}
\end{table}

\subsubsection{Cost-to-Revenue Ratio of Arbitrageurs}
\begin{table}[t!] \centering
\caption{
 Results from regressing the cost-to-revenue ratio of arbitrages on whether the dark venue is introduced and whether the arbitrage order is sent through the dark venue.  The data for regression  covers a sample period from May 4, 2020 to Jul 31, 2021.  Asterisks denote significance levels (***=1\%, **=5\%, *=10\%).} 
\label{regression: costrev}
\begin{tabular*}{0.8\textwidth}{@{\extracolsep{\fill}}  c c c}
\\[-1.8ex]\hline
\hline \\[-1.8ex]
& \multicolumn{2}{c}{\textit{Dependent variables: Cost-to-revenue Ratio}} \
\cr \cline{2-3}
\\  &  (a) &   (b) \\
\hline \\[-1.8ex]
 Intercept & 0.300$^{***}$ & 0.300$^{***}$ \\
  & (0.001) & (0.001) \\
    After & 0.091$^{***}$ & 0.013$^{***}$ \\
  & (0.001) & (0.001) \\ 
  Dark &  & 0.441$^{***}$ \\
  &  & (0.002) \\ 
\hline \\[-1.8ex]
 Observations & 428,685	 & 428,685\\
 $R^2$ & 0.03 & 0.19 \\
\hline
\hline \\[-1.8ex]
\textit{Note:} & \multicolumn{2}{r}{$^{*}$p$<$0.1; $^{**}$p$<$0.05; $^{***}$p$<$0.01} \\
\end{tabular*}
\end{table}

We estimate the following linear models to compare cost-to-revenue ratio of arbitrageurs before and after the introduction of the dark venue:
\begin{equation}\label{regression2}
    CostRevRatio =  \rho_2 \ind_{After}  + \epsilon,
\end{equation}
\begin{equation}\label{regression3}
    CostRevRatio =  \rho_3 \ind_{After} + \rho_4 \ind_{Dark} + \epsilon,
\end{equation}
where $CostRevRatio$ is the cost-to-revenue ratio of arbitrage transaction{s}, $\ind_{After}$ is a dummy variable for the period after the introduction of the dark venue,  $\ind_{Dark}$ is a dummy variable for transaction submitted through dark venue, and $\epsilon$ is an error term.  The coefficient $\rho_2$ quantifies the difference in cost-to-revenue ratio of arbitrages before and after the introduction of the dark venue. The coefficient $\rho_4$ quantifies the difference between the cost-to-revenue ratio of arbitrages sent through the lit venue and arbitrages sent through the dar venue, after the introduction of the dark venue. 

Table \ref{regression: costrev} (a) indicates that,  
after the introduction of the dark venue, the average cost-to-revenue ratio of arbitrageurs increases by around 0.09, a increment that is almost a third of the average cost-to-revenue ratio before the introduction of the dark venue (around 0.3).    Table \ref{regression: costrev} (b) indicates that the average cost-to-revenue ratio of arbitrageurs in the dark venue is 0.44 higher than that of arbitrageur using the lit venue. This suggests that the increase in the cost-to-revenue ratio after the introduction of the dark venue can be mostly attributed to arbitrageurs who use the dark venue.  All results are statistically and  economically significant.  The regression results support our model prediction {that the introduction of the dark venue increases the cost of arbitrageurs and lowers their welfare. }


\subsubsection{The Migration of Users}
\begin{table}[t!] \centering
\caption{
 Results from regressing the proportion of frontrunnable transaction sent through dark venue on the probability of being frontrun. The data for regression covers a sample period from Feb 11, 2020 to May 1, 2021. Asterisks denote significance levels (***=1\%, **=5\%, *=10\%).} \label{regression: user}
\begin{tabular*}{0.8\textwidth}{@{\extracolsep{\fill}}  c c }
\\[-1.8ex]\hline
\hline \\[-1.8ex]
& \textit{Dependent variables: } \\
&  \textit{Proportion of Transactions Through Dark Venue} \\
\hline \\[-1.8ex]
 Intercept & -0.066$^{**}$  \\
  & (0.18) \\
  Probability of Being Frontrun  & 0.605$^{***}$ \\
   & (0.010) \\ 
\hline \\[-1.8ex]
 Observations 	 & 80\\
 $R^2$ & 0.3 \\
\hline
\hline \\[-1.8ex]
\textit{Note:} & {$^{*}$p$<$0.1; $^{**}$p$<$0.05; $^{***}$p$<$0.01} \\
\end{tabular*}
\end{table}
We estimate the following linear model to measure the relationship between users' probability of being frontrun and their venue choice:
\begin{equation}\label{regression4}
    ProportionDark =   \kappa FrontrunProb + \epsilon ,
\end{equation}
$ProportionDark$ is the proportion of frontrunnable transactions sent through the dark venue, $FrontrunProb$ is the probability of being frontrun for transactions sent through the lit venue, and $\epsilon$ is an error term. The coefficient $\kappa$ quantifies the sensitivity of  users' venue selection on the frontrunning risk faced by users. 

Table \ref{regression: user}  indicates that an increase in the probability of being frontrun is positively correlated (60\% correlation) with a higher proportion of transactions sent through the dark venue. A 1\% increase in probability of being frontrun is associated with a 0.6\% increase in the proportion of frontrunnable transactions submitted through the dark venue. The coefficient estimates indicate that these relationships are statistically and  economically significant. 
In summary, Table \ref{regression: user} supports our model prediction {that frontrunnable users migrate from the lit venue to the dark venue} when they face higher frontrun risk.

\newpage
\bibliographystyle{ACM-Reference-Format}
\bibliography{sample-bibliography}

\appendix
\newpage
\section{Technical Results and Proofs}

\begin{proof}[Proofs of Proposition~\ref{venue choice arbitrageurs}, \ref{strategy arbitrageurs}]

We first outline all six potential equilibrium outcomes for venue selection of arbitrageurs. We then solve for the equilibrium transaction fee bidding strategies in all six cases. Finally, we solve for the equilibrium venue selection strategies of arbitrageurs. 

There are six potential equilibrium outcomes for arbitrageurs' venue selection: (1) Both arbitrageurs choose the dark venue; (2) One arbitrageur chooses the dark venue, and the other arbitrageur chooses the lit venue; (3) One arbitrageur chooses the dark venue, and the other arbitrageur chooses both venues; (4) One arbitrageur chooses the lit venue, and the other arbitrageur chooses both venues; (5) Both arbitrageurs choose the lit venue; (6) Both arbitrageurs choose both venues. 

\paragraph{Case 1: Both arbitrageurs choose the dark venue. }
We show that there is no pure strategy Nash equilibrium (PNE), and there exists a unique mixed strategy Nash equilibrium (MNE) where both arbitrageurs bid $g\in[v_{B-2}, c]$, and $g$ follows the probability distribution 

$$P(g)=\left\{\begin{matrix}
\frac{1-p}{p}\cdot\frac{1}{(1-\frac{g-v_{B-2}}{c-v_{B-2}})^2\cdot (c-v_{B-2})} & v\leq (c-v_{B-2})\cdot p+v_{B-2}\\ 
0 & v>(c-v_{B-2})\cdot p+v_{B-2}
\end{matrix}\right.$$

We prove the non-existence of PNE  in two steps. First, we show that there is no symmetric PNE using a contradiction argument. Second, we show that there is no asymmetric PNE.

Assume there is a symmetric PNE where both arbitrageurs bid the same transaction price $f_{D_i} = f_{D_j} = g$, and the expected utility of arbitrageur $i$ is not higher than the expected utility of arbitrageur $j$. We argue that there exists an unilateral deviation which allows arbitrageur $i$ to improve its expected utility. 
If $g<c$, the expected utility of arbitrageur $i$ is $A_i \leq (1-p)\cdot(c-g)+\frac{p}{2}\cdot(c-g_i)$. Arbitrageur $i$ can increase its expected utility by changing its strategy to  $f_{D_i}' = g + \epsilon$. Its expected payoff would then be $A_i' = c-(g+\epsilon) > A_i$.
If $g=c$, the expected utility of arbitrageur $i$ is 0. Arbitrageur $i$ can then deviate to a strategy $f_{D_i}'  = v_{B-2}$. Then its expected payoff is $A_i' = p\cdot(c-v_{B-2}) > 0$. Therefore, there exists no symmetric PNE.

We next argue that there exists no asymmetric PNE. Assume there exists a PNE where $f_{D_i} < f_{D_j}$. We argue that one of the bidding arbitrageurs can improve its expected utility by deviating its strategy.
If $f_{D_i} = g > v_{B-2}$, the expected utility of arbitrageur $i$ is $A_i = p\cdot(c-g)$. Therefore, arbitrageur $i$ can deviate to a strategy with  $f_{D_i}' = v_{B-2}$. In such a case, $A_i' = p\cdot(c-v_{B-2}) > A_i$.
If $f_{D_i} = v_{B-2}$ and $f_{D_j} = g > v_{B-2}+\epsilon$, the expected utility of arbitrageur $j$ is $A_j = c-g$. Therefore, arbitrageur $j$ can deviate to a strategy where $f_{D_j}' = v_{B-2}+\epsilon$. In this case, $A_j' = c-(v_{B-2}+\epsilon) > A_j$.
If $f_{D_i} = v_{B-2}$ and $f_{D_j} = v_{B-2}+\epsilon$, the expected utility of arbitrageur $i$ is $A_i = p\cdot(c-v_{B-2})$. Therefore, arbitrageur $i$ can deviate to a strategy with $f_{D_j}' = v_{B-2}+2\epsilon$. In this case, $A_i' = c-(v_{B-2}+2\epsilon) > A_i$.  Therefore, there exists no asymmetric PNE.

Next, we discuss MNE. We show that there exists no pure strategy which yield a higher expected utility than the mixed strategy for all players.

When arbitrageur $i$ play the mixed strategy, its expected utility is

\begin{small}
\begin{align*}
    &A_i = (1-p)\cdot\int _{v_{B-2}}^{(c-v_{B-2})\cdot p+v_{B-2}}P(t)\cdot(c-t)\mathrm{d}t\\&+p\cdot\int _{v_{B-2}}^{(c-v_{B-2})\cdot p+v_{B-2}}P(t)\cdot(\int _{v_{B-2}}^t P(s)\mathrm{d}s)\cdot(c-t)\mathrm{d}t\\ 
    &= (1-p)\cdot\int _{v_{B-2}}^{(c-v_{B-2})\cdot p+v_{B-2}}\frac{1-p}{p}\cdot\frac{1}{(1-\frac{t-v_{B-2}}{c-v_{B-2}})^2\cdot (c-v_{B-2})}\cdot(c-t)\mathrm{d}t\\&+p\cdot\int _{v_{B-2}}^{(c-v_{B-2})\cdot p+v_{B-2}}\frac{1-p}{p}\cdot\frac{1}{(1-\frac{t-v_{B-2}}{c-v_{B-2}})^2\cdot (c-v_{B-2})}\cdot\frac{(1-p)\cdot(t-v_{B-2})}{p\cdot(c-t)}\cdot(c-t)\mathrm{d}t\\ 
    &=(1-p)\cdot(c-v_{B-2})
\end{align*}
\end{small}

Then we show that the other bidding strategy cannot outperform the MNE strategy. We first consider the pure strategy where $f_{D_i}'=  >(c-v_{B-2})\cdot p+v_{B-2}$. The bidder will then always win the game. Therefore, $A_j'=c - -f_{D_i}'<(1-p)\cdot(c-v_{B-2})$, which indicates that bidders are not better off deviating.

Next, we consider the pure strategy where $f_{D_i}' \leq (c-v_{B-2})\cdot p+v_{B-2}$. We can write the expected utility of  arbitrageur $i$ as
\begin{align*}
    &A_i' &=(1-p)\cdot(c-f_{D_i}')+p\cdot(c-f_{D_i}')\cdot\int _{v_{B-2}}^{f_{D_i}'}P(t)\mathrm{d}t= (1-p)\cdot(c-v_{B-2}).
\end{align*}

Therefore, deviating to another strategy $f_{D_i}'$ cannot increase the expected utility of arbitrageur $i$ when the other bidder plays the mixed strategy. Therefore, a combination of any pure strategies cannot outperform the mixed strategy.

\paragraph{Case 2: one arbitrageur chooses the dark venue, and the other arbitrageur chooses the lit venue.}
As there is no competition in the dark venue, the arbitrageur in the dark venue will bid the lowest bid $v_{B-1}$ when he observes an arbitrage opportunity or finds the other arbitrageur's bid in the lit venue. The arbitrageur in the lit venue also bids $v_{B-1}$ because he is the only bidder in the lit venue.

\paragraph{Case 3: one arbitrageur chooses he dark venue, and the other arbitrageur chooses both venues. }
The arbitrageur acting in both venues bids $c$ in the dark venue and $v_{B-2}$ in the lit venue. It knows that this information will be leaked to the other arbitrageur. It bids the lowest bid in the lit venue as there is no competition. It bids truthfully in the dark venue because this is a sealed-bid first-price auction, where both bidders have the same valuation $c$. 
The arbitrageur acting only in the dark venue observes the other arbitrageur bidding in the lit venue. Then it will bid $c$ in the dark venue. This is because in the dark venue, the bidding mechanism is a sealed-bid first-price auction where both bidders have the same valuation. If the arbitrageur finds an opportunity, and it does not observe the bid of the other arbitrageur, it will just bid $v_{B-2}$ because there is no competition. 

\paragraph{Case 4: one arbitrageur chooses the lit venue, and the other arbitrageur chooses both venues.}
We first consider the arbitrageur which submits too both venue. This arbitrageur always bids $v_{B-2}$ in the dark venue because there is no competition in the dark venue. We then consider both  arbitrageurs' strategies in the lit venue. It is obvious that the one which submits the opening bid will be $v_{B-1}$ because this lowers its transaction cost. If the auction ends in this round, then its transaction cost is minimized. If the auction continues, setting up an opening bid as $v_{B-1}$ also lowers its expected transaction cost. For each round, both arbitrageurs just increase by the minimal increment $\epsilon$ because this lowers their transaction cost.

\paragraph{Case 5: both arbitrageurs choose the lit venue.}
If both arbitrageurs choose the lit venue, their bidding strategy  in the lit venue is exactly the same as their strategy in Case 4.

\paragraph{Case 6: both arbitrageurs choose both venues} If both arbitrageurs choose both venues, they all bid truthfully in the dark venue. This is because the bidding mechanism is a sealed-bid, first-price auction where both arbitrageurs have the same valuation. In the lit venue, they all use the same bidding strategy as in Case 4. 

We then calculate the expected equilibrium payoff of each arbitrageur in all six cases, and construct the following matrix:

\begin{table}[h!]
\footnotesize
\setlength\tabcolsep{1.5pt} 
\begin{tabular}{|l|l|l|l|}
\hline
$\begin{matrix}
           A_{1} ,\\
           A_{2}
         \end{matrix}$ & Dark & Lit & All \\ \hline
Dark    & $\begin{matrix}
          \alpha p(1-p)(c-v_{B-2}), \\
          \alpha p(1-p)(c-v_{B-2})
         \end{matrix}$  &  $\begin{matrix}
          \alpha (1-(1-p)^2)(c-v_{B-2}), \\
           (1-\alpha) p (c-v_{B-2})
         \end{matrix}$   &  $\begin{matrix}
          \alpha p(1-p)(c-v_{B-2}), \\
           (1-\alpha) p (c-v_{B-2})
         \end{matrix}$   \\ \hline
Lit     &   $\begin{matrix}
          (1-\alpha) p (c-v_{B-2}), \\
           \alpha (1-(1-p)^2)(c-v_{B-2})
         \end{matrix}$     &  $\begin{matrix}
          \frac{1}{2}(c-\gamma v_{B-2})(1-(1-p)^2), \\
         \frac{1}{2}(c-\gamma v_{B-2})(1-(1-p)^2)
         \end{matrix}$       &  $\begin{matrix}
          \frac{1}{2} (1-\alpha) (c-\gamma v_{B-2})(1-(1-p)^2), \\
        ( \frac{1}{2}(c-\gamma v_{B-2}) (1-\alpha) \\ + \alpha( c-v_{B-2}) (1-(1-p)^2) 
         \end{matrix}$     \\ \hline
All     & $\begin{matrix}
(1-\alpha) p (c-v_{B-2}),\\
          \alpha p(1-p)(c-v_{B-2}) 
         \end{matrix}$        &   $\begin{matrix}  
        ( \frac{1}{2}(c-\gamma v_{B-2}) (1-\alpha) \\ + \alpha( c-v_{B-2}) (1-(1-p)^2) , \\
          \frac{1}{2} (1-\alpha) (c-\gamma v_{B-2})(1-(1-p)^2) 
         \end{matrix}$  &    $\begin{matrix}   ( \frac{1}{2}(c-\gamma v_{B-2}) (1-\alpha) ) (1-(1-p)^2) ,\\
           ( \frac{1}{2}(c-\gamma v_{B-2}) (1-\alpha) ) (1-(1-p)^2)
         \end{matrix}$  \\ \hline
\end{tabular}
\end{table}
where $\gamma >1 , \gamma v_{B-2} <c$. 

We next solve for the equilibrium venue selection strategy of arbitrageurs. 

If $\alpha > \alpha_2 = \frac{1}{2-p}$, $\alpha p(1-p)(c-v_{B-2}) > ( \frac{1}{2}(c-\gamma v_{B-2}) (1-\alpha) ) (1-(1-p)^2) $, and $\alpha p(1-p)(c-v_{B-2}) > (1-\alpha) p (c-v_{B-2}) $. Those two conditions ensure that the unique equilibrium is that both arbitrageurs choose the dark venue. 

If $\alpha < \alpha_1 = \frac{p \gamma -2 \gamma }{p \gamma +p-2 \gamma -1}$, $\alpha p(1-p)(c-v_{B-2}) < (1-\alpha) p (c-v_{B-2}) $ and $\alpha p(1-p)(c-v_{B-2}) > ( \frac{1}{2}(c-\gamma v_{B-2}) (1-\alpha) ) (1-(1-p)^2) $. Using the tie-break rule and the two conditions above, the unique equilibrium is that both arbitrageurs choose both venues. 

If $\alpha_2> \alpha > \alpha_1 $, we have $\alpha p(1-p)(c-v_{B-2}) < ( \frac{1}{2}(c-\gamma v_{B-2}) (1-\alpha) ) (1-(1-p)^2) $, and $\alpha p(1-p)(c-v_{B-2}) > (1-\alpha) p (c-v_{B-2}) $. Those two conditions ensure that one arbitrageur choosing both venues, and the other arbitrageur choosing the dark venue is the equilibrium.

\end{proof}

\begin{proof}[Proof of Proposition~\ref{users proposition}]
 We only prove the proposition in the case $\alpha \in (\alpha_2,1]$. The other two cases can be shown using the same procedure.  If $\alpha \in (\alpha_2,1]$, by Proposition \ref{venue choice arbitrageurs}, both arbitrageurs choose the dark venue. 
 
 If the frontrunnable user chooses the dark venue, her expected payoff is $$\alpha (v_0-v_{B-1}).$$
 
 If instead the frontrunnable user chooses the lit venue, her expected payoff is $$((1-\alpha)+\alpha(1-p)^2) (v_0-v_{B-2})+ \alpha(1-(1-p)^2) (v_0-c - v_{B-2})).$$

 Comparing the payoff in the two venues, we have that the frontrunnable user chooses the dark venue if and only if $ \alpha > \lambda_1 = \frac{v_0-v_{B-1}}{-c p^2+2 c p+v_{B-1} p^2-2 v_{B-1} p-v_{B-1}-v_{B-2} p^2+2 v_{B-2} p+v_0} $
 \end{proof}
 
 \begin{proof}[Proof of Proposition~\ref{equilibrium: benchmark}]
  Suppose $\alpha = 0$. If the frontrunnable user submits to the lit venue, then the payoff of the frontrunnable user is $$(1-p)^2 (v_0-v_{B-2})+ (1-(1-p)^2) (v_0 -c -v_{B-2})).$$ 
  
  The quantity above is positive if and only if $c < c_1 = \frac{v_0-v_{B-2}}{(1-(1-p)^2) }$. If it is positive, then the frontrunnable user will submit her transaction. Otherwise, she will not submit to the blockchain. 
  
 \end{proof}
 
 \begin{proof}[Proof of Proposition \ref{equilibrium: from lit to dark}]
 If $c > c_1$, the frontrunnable user will only use the dark venue. This is because using the lit venue generates a payoff $(1-p)^2 (v_0-v_{B-2})+ (1-(1-p)^2) (v_0 -c -v_{B-2})) < 0$, while using the dark venue generates a payoff $\alpha(v_0 -v_{B-2}) \geq 0$
 
 Miners in the lit venue earn $r_{lit}(\alpha) =  B v_{B+1}$ after mining a block. For any sufficiently small mass $\delta > 0$ of  miners who migrate from the lit to the dark venue, they earn $r_{dark}(\alpha + \delta) = Bv_{B} > Bv_{B+1}$. In equilibrium, all miners adopt the dark venue. 
 
 If $c \leq c_1$, we can show that $\lambda_1$ is a equilibrium, and it is easy to verify that the other equilibria are $\lambda_2, \lambda_3, 1$.
 
 At $\lambda_1$, for a sufficiently small mass $\delta > 0$ of  miners migrating to the dark venue, their payoffs in the dark venue are equal to  $(B-1) v_{B-2} +(1-p)^2v_{B-1} + c(1-(1-p)^2)$. If they migrate to the lit venue,  their payoff in the lit venue are $(B-1) v_{B-2} +(1-p)^2v_{B-1} +(1-(1-p)^2) \gamma v_{B-2} <(B-1) v_{B-2} +(1-p)^2v_{B-1} + c(1-(1-p)^2)$. Hence, there is no incentive for them to migrate. For a sufficiently small mass $\delta > 0$ of  miners in lit venue, the payoff is equal to $(B-1) v_{B-2} +(1-p)^2v_{B-1} +(1-(1-p)^2) \gamma v_{B-2}$. If they migrate to the dark venue, their payoff is equal to $r_{dark}(\lambda_1 + \delta) = B v_{B-1} < (B-1) v_{B-2} +(1-p)^2v_{B-1} +(1-(1-p)^2) \gamma v_{B-2}.$ There is no incentive for them to migrate. This is because if $\alpha> \lambda_1$, the frontrunnable user migrates to the dark venue, and there is no longer a frontrunning arbitrage. At $\lambda_1$, the frontrunnable user still submits to the lit venue as shown in Proposition~\ref{users proposition}, and the arbitrageurs submit to both venues as shown in Proposition~\ref{venue choice arbitrageurs}. 
 \end{proof}
 
 \begin{proof}[Proof of Proposition \ref{transaction cost}]
  
  If $c>c_1$, frontrunnable trader does not submit transactions in the mempool, thus the minimum fee that guarantees the execution of a transaction is $v_B$ and the total fee of all transactions are $B\cdot v_B$. With the introduction of a dark venue, the execution fee increases to $v_{B-1}$, while the total fee increases to $B\cdot v_{B-1}$.
  
  If $c\leq c_1$, the minimum fee that guarantees the execution of a transaction is always $v_{B-2}$. The expected total fee of all transactions before the introduction of a dark venue is $v_{B-2}*(B-1)+(1-p)^2v_{B-1}+(1-(1-p)^2)\gamma v_{B-2}$, while the expected fee increases to  $v_{B-2}*(B-1)+(1-p)^2v_{B-1}+(1-(1-p)^2)c-2p(1-p)(c-v_{B-2})$ and $v_{B-2}*(B-1)+(1-p)^2v_{B-1}+(1-(1-p)^2)(c-v_{B-2})$ in partially adoption Nash equilibria.
 
 \end{proof}
 
 \begin{proof}[Proof of Proposition \ref{welfare analysis}]
 
 We compare the welfare of miners, frontrunnable users, and arbitrageurs separately before and after the introduction of the dark venue.
 
 Before the introduction of the dark venue, with probability $1-(1-p)^2$, the transaction of frontrunnable user will be observed by arbitrageurs. Therefore, the expected payoff of the frontrunnable user before the introduction of the dark venue is
 
 $$v_0-v_{B-2}-(1-(1-p)^2)c$$.
 
 The expected payoff of the winning miner is
 
 $$v_{B-2}*(B-1)+(1-p)^2v_{B-1}+(1-(1-p)^2)\gamma v_{B-2}$$.
 
 The expected payoff of arbitrageurs is
 
 $$\frac{1}{2}(c-\gamma v_{B-2})(1-(1-p)^2)$$.
 
 The expected payoff of all non-frontrunnable users is
 
 $$\sum_{i=1}^{B-2}v_i-v_{B-2}*(B-2)$$.
 
 Then, we consider the welfare of different stakeholders in Nash equilibria.
 
 
 In the first Nash equilibrium, the frontrunnable user selects the lit venue and the arbitrageurs select both the dark venue. The expected payoff of the frontrunnable user is $v_0-v_{B-2}-(1-(1-p)^2)c$. 
 The expected payoff of the winning miner if she joins the dark venue is $v_{B-2}*(B-1)+(1-p)^2v_{B-1}+(1-(1-p)^2)c-2p(1-p)(c-v_{B-2})$.
  The expected payoff of the winning miner if she stays in the lit venue is $v_{B-2}*(B-1)+v_{B-1}$.
  The payoff of arbitrageurs is $\alpha p (1-p)(c-v_{B-2})$. The expected payoff of all non-frontrunnable users is $\sum_{i=1}^{B-2}v_i-v_{B-2}*(B-2)$.
 
 In the second Nash equilibrium, the frontrunnable user selects the lit venue and the arbitrageurs select both venues. The expected payoff of the frontrunnable user is $v_0-v_{B-2}-(1-(1-p)^2)c$. The payoff of the winning miner if she joins the dark venue is $v_{B-2}*(B-1)+(1-p)^2v_{B-1}+(1-(1-p)^2)c$. The expected payoff of the winning miner if she stays in the lit venue is $v_{B-2}*(B-1)+(1-p)^2v_{B-1}+(1-(1-p)^2)\gamma v_{B-2}$.
  The payoff of arbitrageurs is $( \frac{1}{2}(c-\gamma v_{B-2}) (1-\alpha) ) (1-(1-p)^2)$. The expected payoff of all non-frontrunnable users is $\sum_{i=1}^{B-2}v_i-v_{B-2}*(B-2)$.
 
 In the third Nash equilibrium, the frontrunnable user selects the lit venue, while one arbitrageur selects both venues. The expected payoff of the frontrunnable user is $v_0-v_{B-2}-(1-(1-p)^2)c$. The payoff of the winning miner if she joins the dark venue is $v_{B-2}*(B-1)+(1-p)^2v_{B-1}+(1-(1-p)^2)(c-v_{B-2})$. The expected payoff of the winning miner if she stays in the lit venue is $v_{B-2}*(B-1)+(1-p)^2v_{B-1}+(1-(1-p)^2)\gamma v_{B-2}$.
  The payoff of arbitrageurs are $( \frac{1}{2}(c-\gamma v_{B-2}) (1-\alpha) ) (1-(1-p)^2)$ and $\frac{1}{2}(c-\gamma v_{B-2})(1-\alpha)+\alpha(c-v_{B-2})(1-(1-p)^2)$.
 
 \end{proof}

  \begin{proof}[Proof of Proposition \ref{aggregate welfare}]
  
  The aggregate welfare of all stakeholders is the sum of the valuations of transactions included in the block.
  
  If $c> c_1$, then the frontrunnable trader does not submit transactions before the introduction of the dark venue. Therefore, the aggregate social welfare of stakeholders is $\sum_{i=1}^{B} v_i$. Because the full adoption is the only equilibrium in this scenario, the aggregate social welfare will increase to $\sum_{i=0}^{B-1} v_i$ after the introduction of the dark venue.
  
  If $c\leq c_1$, the expected aggregate social welfare of stakeholders before the introduction of the dark venue is $(1-p)^2\sum_{i=0}^{B-2} v_i + (1-(1-p)^2)\sum_{i=0}^{B-1} v_i$. 
  
  The expected aggregate social welfare of stakeholders after the introduction of the dark venue is
  
  \begin{itemize}
      \item $(1-\alpha) (\sum_{i=0}^{B-1} v_i)+\alpha\cdot((1-p)^2\sum_{i=0}^{B-2} v_i + (1-(1-p)^2)\sum_{i=0}^{B-1} v_i)$, if both arbitrageurs select the dark venue;
      \item $(1-p)^2\sum_{i=0}^{B-2} v_i + (1-(1-p)^2)\sum_{i=0}^{B-1} v_i$, if both arbitrageurs select both venues;
      \item $(1-p)^2\sum_{i=0}^{B-2} v_i + (1-(1-p)^2)\sum_{i=0}^{B-1} v_i$, if one arbitrageur selects the lit venue and the other selects both venues.
  \end{itemize}
  
  Therefore, the introduction of the dark venue weakly raises aggregate welfare in all Nash equilibria.
  
  If the dark venue is fully adopted, then the sum of the valuations of transactions included in the block is $\sum_{i=0}^{B-1} v_i$. If the dark venue is only partially, arbitrage transactions might be included in the block if the winning miner joins the dark venue. As arbitrage transactions does not generate social welfare and have substituted another non-frontrunnable transaction. Therefore, the largest expected aggregate social welfare in all NE is $(1-\alpha) (\sum_{i=0}^{B-1} v_i)+\alpha\cdot((1-p)^2\sum_{i=0}^{B-2} v_i + (1-(1-p)^2)\sum_{i=0}^{B-1} v_i)< \sum_{i=0}^{B-1} v_i$.
 
 \end{proof}

  
  
  
  
 

   \begin{proof}[Proof of Proposition \ref{transfer design}]
    If $c> c_1$, there exists a unique full adoption equilibrium at which the aggregate welfare is maximized. The required payment is then zero.

    If $c\leq c_1$, then there exists a partial adoption equilibrium. At the partial adoption equilibrium, the adoption rate of the dark venue is $\alpha^* \in \{\lambda_1, \lambda_2, \lambda_3\}$. We only prove the case where $\alpha^* = \lambda_1$, and the other two cases can be shown with the same procedure. At equilibrium, the expected arbitrage loss of the frontrunnable user is $(1-(1-p)^2)c$. $(1-(1-p)^2)c$ is also the sum of expected arbitrage revenue of two arbitrageurs. The sum of expected transaction fees paid by two arbitrageurs is $(1-(1-p)^2) \gamma v_{B-2}$, where $c > \gamma v_{B-2}$ as the arbitrageurs extract non-negative profit from frontrunning. Assume that frontrunnable user commits to pay $(1-(1-p)^2)c$ to the winning miner who has adopted the dark venue. When $v_{B-2} - v_{B-1} $ is sufficiently small, $r_{dark}(\lambda_1 + \delta) = B v_{B-1} + (1-(1-p)^2)c > (B-1) v_{B-2} +(1-p)^2v_{B-1} +(1-(1-p)^2) \gamma v_{B-2} = r_{lit}(\lambda_1 + \delta).$ In this way, a marginal miner will migrate to the dark venue, and any partial adoption equilibrium does not exist. Besides, the frontrunnable user is not worse off after making the payment.

 \end{proof}

\newpage
\section{Empirical Methodology}

\subsection{Frontrunning Arbitrages}

\label{sec:fron1}

In this section, we explain the methodology used to identify frontrunning arbitrages. We identify a two-legged trade ($T_{A1}$, $T_{A2}$) as a frontrunning arbitrage, and a transaction $T_V$ as the corresponding victim transaction, if the following conditions are met:

\begin{enumerate}
\item $T_{A1}$ and $T_{A2}$ are included in the same block, and $T_{A1}$ is executed before $T_{A2}$. $T_{A1}$ and $T_{A2}$ have different transaction hashes.
\item $T_{A1}$ and $T_{A2}$ swap assets in the same liquidity pool, but in opposite directions. The input amount for the swap in $T_{A2}$ is equal to the output amount of the swap in $T_{A1}$. In this way, the transaction $T_{A2}$ closes the position built up in the first leg {$T_{A1}$}.  
\item $T_V$ is executed between $T_{A1}$ and $T_{A2}$. $T_V$ swaps assets in the same liquidity pool as $T_{A1}$ and $T_{A2}$. $T_V$ swaps assets in the same direction as $T_{A1}$.
\item Every transaction $T_{A2}$ is mapped to exactly one transaction $T_{A1}$.
\end{enumerate}

There exists frontrunning arbitrages where $T_{A1}$ and $T_{A2}$ are placed in different blocks. However, arbitrageurs normally prefer to include $T_{A1}$ and $T_{A2}$ in one block to minimize inventory risk. Nonetheless, the above procedure allows us to find a lower bound for the number of frontrunning arbitrages. The revenue  of a frontrunning arbitrage is the difference between the output of $T_{A2}$ and the input of $T_{A1}$, and the profit is the revenue minus the gas fee paid for these two transactions.




\subsection{Frontrunnable Transactions}
In this section, we provide that methodology to identify transactions vulnerable to frontrunning arbitrages. Observe that not all frontrunnable transactions are exploited by arbitrageurs. 

\label{sec:fron2}

There were~$17,644,672$ transactions in the given time frame. The input token {of $9,003,759$ of these transactions} is ETH. We only focus on those transactions. This is because most arbitrageurs are bots, and only conduct arbitrages where ETH serves as input token.  For each transaction, we calculate the optimal revenue that an arbitrageur can attain by frontrunning this transaction. If the revenue is positive, then we identify this transaction as  frontrunnable.

A swap transaction often has a slippage tolerance threshold $m$ which specifies the minimum amount of output token to be received in the transaction.  If the price impact of the frontrunning transaction $T_{A1}$ is too large, the slippage tolerance threshold of the victim transaction $T_V$ may be triggered and $T_V$ will automatically fail. In this case, the arbitrage will not be profitable. This is why we have to account for the slippage tolerance threshold for each swap transaction in our calculation. Formally, let $v$ be the amount of input token specified in the victim transaction $T_V$, and $m$ the minimum amount of output token to be received. Let $x$ be the amount of input token swapped in the frontrunning transaction $T_{A1}$. 
Let $r_1$ and $r_2$ represent the liquidity reserves of input token and output token in the pool. The transaction fee in Uniswap and Sushiswap is 0.3\%.  The victim transaction will not fail if 

$$ \frac{v \cdot 0.997 \cdot (r_2 - \frac{x \cdot 0.997 \cdot r_2}{r_1 + 0.997\cdot x})}{(r_1 + x) + 0.997\cdot v} \geq m. $$

We solve the largest $x$ that satisfies the above inequality. The result can be written   as
$$ maxInput_{A1}(r_1, r_2, v, m) = \frac{5.01505 \cdot 10^{-7} \cdot t}{\sqrt{m}} - 1.0015 r_1 - 0.4985 v, $$
where
$$ t = \sqrt{9000000 r_1^2 m + 3976036000000 r_1 r_2 v - 5964054000 r_1 m v + 988053892081 m v^2}.$$

The $maxInput_{A_1}$  is  the largest trade size  of transaction $T_{A1}$ such that $T_V$ will not fail. We can then calculate the output amount in the second leg of the arbitrage $T_{A2}$ which closes the position built up in $T_{A1}$.  $T_V$ is frontrunnable if the constructed frontrunning arbitrage yields a positive revenue. {In total, we identify~$3,612,343$}  frontrunnable transactions with ETH as the input token.

\end{document}